\documentclass[]{interact}
\usepackage{amsmath}
\usepackage{mathtools}
\usepackage{geometry}
\usepackage{graphicx}
\usepackage{multirow}
\usepackage{setspace}
\usepackage[utf8]{inputenc}

\graphicspath{    {converted_graphics/}    {/}}

\begin{document}

\title{Thermodynamic properties of the 3D Lennard-Jones/spline model}

\author{\name{Bjørn Hafskjold\textsuperscript{a},\thanks{CONTACT: B. Hafskjold. Email: bjorn.hafskjold@ntnu.no} 
Karl Patrick Travis,\textsuperscript{b} Amanda Bailey Hass,\textsuperscript{b} \\ Morten Hammer,\textsuperscript{c}, Ailo Aasen,\textsuperscript{c,d} and \O{ivind} Wilhelmsen\textsuperscript{c, d}}
\affil{\textsuperscript{a} Norwegian University of Science and Technology, Department of Chemistry, NO-7491 Trondheim, Norway \\
\textsuperscript{b} University of Sheffield, Department of Materials Science and Engineering, Sheffield, UK. \\
\textsuperscript{c} SINTEF Energy Research, NO-7465 Trondheim, Norway \\ \textsuperscript{d} Norwegian University of Science and Technology, Department of Energy and Process Engineering, NO-7491 Trondheim, Norway}
}

\maketitle

\begin{abstract}
In the Lennard-Jones spline (LJ/s) model, the Lennard-Jones (LJ) potential is truncated and splined so that both the pair potential and the force continuously approach zero at $r_c \approx 1.74 \sigma$. It exhibits the same structural features as the LJ model, but the thermodynamic properties are different due to the shorter range of the potential. One advantage of the model is that simulation times are much shorter. In this work, we present a systematic map of the thermodynamic properties of the LJ/s model from molecular dynamics and Gibbs ensemble Monte Carlo simulations. Accurate results are presented for gas/liquid, liquid/solid and gas/solid coexistence curves, supercritical isotherms up to a reduced temperature of 2 (in LJ units), surface tensions, speed of sound, the Joule-Thomson inversion curve, and the second to fourth virial coefficients. The critical point for the LJ/s model is estimated to be $T_\text{c}^*=0.885 \pm 0.002$ and $P_\text{c}^*=0.075 \pm 0.001$, respectively. The triple point is estimated to $T_\text{tp}^*=0.547 \pm 0.005$ and $P_\text{tp}^*=0.0016 \pm 0.0002$. Despite the simplicity of the model, the acentric factor was found to be as large as $0.07 \pm 0.02$. The coexistence densities, saturation pressure, and supercritical isotherms of the LJ/s model were fairly well represented by the Peng-Robison equation of state. We find that Barker-Henderson perturbation theory works much more poorly for the LJ/s model than for the LJ model. The first-order perturbation theory overestimates the critical temperature and pressure by about 10\% and 90\%, respectively. A second-order perturbation theory that uses the mean compressibility approximation shifts the critical point closer to simulation data, but makes the prediction of the saturation densities worse. It is hypothesized that the reason for this is that the mean compressibility approximation gives a poor representation of the second-order perturbation term for the LJ/s model and that a correction factor is needed at high densities. Our main conclusion is that we at the moment do not have a theory or model that adequately represents the thermodynamic properties of the LJ/s system.
\end{abstract}

\maketitle
\begin{doublespacing}

\section{Introduction}
The Lennard-Jones (LJ) potential is a simple model capable of describing many real systems, in particular the noble gases. The model’s structural, thermodynamic, and transport properties have been extensively examined and documented, see \textit{e.g.} Rutkai \textit{et al.} \cite{rutkai2017}. When used in computer simulations, the intermolecular potential is truncated, typically at a distance of $2.5 - 5$ molecular diameters. This truncation has little effect on the system’s structure, but in order to fully represent the LJ system's thermodynamic properties, the contributions from the long-range tail must be included by use of tail corrections~\cite{nicolas1979}. Some properties such as the surface tension, are especially sensitive to truncation and whether the potential also has been shifted. Trokhymchuk and Alejandre argue that truncation distances exceeding 4 in reduced units in addition to a long-range correction was necessary to accurately represent the surface tension of the full LJ model~\cite{trokhymchuk1999}. Hence, when results of a simulation study are reported, care must be taken to specify exactly how the potential is truncated and corrected.

A convenient alternative to the LJ potential is the Lennard-Jones spline (LJ/s) potential introduced by Holian and Evans \cite{holian1983}. The LJ/s potential is a LJ potential truncated in a unique way. As such, it avoids the need for further specification and risk of ambiguity in how the potential is used in a simulation. The pair potential of the LJ/s model is 
\begin{equation}
u(r)=
  \begin{cases}
  4\varepsilon\left[ \left( \frac{\sigma}{r}\right)^{12}-\left( \frac{\sigma}{r}\right)^{6}  \right]  & \text{for }r<r_s, \\ 
  a(r-r_c)^2+b(r-r_c)^3 & \text{for }r_{s}<r<r_c,  \\ 
  0 & \text{for }r>r_c,
  \end{cases}
  \label{eqn:ljs}
\end{equation}
where $\varepsilon$ and $\sigma$ are the usual Lennard-Jones parameters.
The distance $r_s$ is given by the LJ potential’s inflection point. The parameters $a$, $b$, and $r_c$ are determined such that the potential and its derivative are continuous at $r_s$ and $r_c$. This means that the force is zero at $r_c$ and the delta-function contribution to the force in the LJ potential at the cut-off is avoided. The spline parameters are $r_s=\left ( \frac{26}{7} \right ) ^{(1/6)} \sigma \approx 1.24 \sigma$, $r_c=\frac{67}{48} r_s \approx 1.74 \sigma$, $a=-\frac{24192}{3211} \frac{\varepsilon}{r_s^2}$, and $b=-\frac{387072}{61009}\frac{\varepsilon}{r_s^3}$. The short range of the LJ/s model leads to simulation times that are reduced by as much as 50\% in the liquid state in comparison with a LJ model truncated at $2.5 \sigma$.

The LJ/s model has essentially the same structural features as the LJ model, but since the potential is of shorter range, the thermodynamic properties are different. Although scattered thermodynamic data have been published for the LJ/s model \cite{halseid1993, rosjorde2000}, it appears that no systematic map of its properties has been reported. The main purpose of this paper is therefore to do so. We have used molecular dynamics (MD) and Gibbs ensemble Monte Carlo (GEMC) simulations as the primary tools to examine the model's properties. A second purpose of the paper is to discuss a limited set of current thermodynamic models and theories in light of the simulation data, \textit{viz.} the virial expansion, the Barker-Henderson second-order perturbation theory \cite{barker1967I, barker1967II}, and cubic equations of state.

An equation of state (EoS) provides an important summary of a system’s thermodynamic properties. Since van der Waals' pioneering work on EoS at the end of the 19th century~\cite{Waals_1873}, the field has developed much and branched into different types of EoS, where each type has its advantages and disadvantages. Cubic EoS, such as Peng--Robinson \cite{Peng_1976} and Soave--Redlich--Kwong~\cite{SRK}, possibly with advanced mixing rules~\cite{Huron1979,Holderbaum1991} are used extensively in process simulations, time-consuming optimisation studies, and solubility predictions~\cite{Aasen2017,Koulocheris2018}. 
The accuracy of cubic EoS are limited by their simple form, and they are known to give poor predictions of the liquid-phase densities.
When higher accuracy is needed, corresponding-state methods like Lee--Kesler \cite{Lee1975}, extended corresponding state methods like SPUNG~\cite{Jorstad_1993}, or multi-parameter EoS
like GERG-2008 \cite{Span_1996} are taken advantage of. The most accurate representation of the LJ model has been achieved by using so-called multi-parameter EoS~\cite{Thol2015}. Even though this type of EoS has unparalleled accuracy in the regions where data are available, they exhibit unphysical behaviour in the two-phase region, which makes them unsuitable for several applications such as prediction of interfacial phenomena with density functional theory~\cite{Wilhelmsen2017}.
Whereas several versions of the EoS exist for the LJ system \cite{Thol2015,Johnson_1993}, an EoS for the LJ/s system has not yet been developed. 

The preferred choice for an accurate representation of fluid properties where the interaction potential is available is arguably the perturbation theory developed by Barker and Henderson (BH) \cite{barker1967I}, which has a firm basis in statistical mechanics. The original second-order theory showed good agreement with simulation data for the LJ and square well fluids \cite{barker1967I,barker1967II}. Critical analyses of the BH theory and the Weeks-Chandler-Anderson (WCA) theory applied to the LJ fluid have recently been published \cite{van2017}. When the first three perturbation terms in the expansion are included, both theories show excellent agreement with simulation results. In fact, it was already shown by Barker and Henderson that a second-order theory gave reasonable results for the LJ fluid \cite{barker1967II}. A third-order perturbation theory was also recently presented for Mie-fluids~\cite{Lafitte2013}. 

A study of the square-well potential showed that BH theory and simulations were in better agreement for longer ranged potentials \cite{solana2013}. It may therefore be expected that the BH theory for the LJ/s model is inferior to that of the LJ model. This hypothesis will be further explored in this paper.

The outline of this paper is as follows. Simulations of $PV$-isotherms, phase diagram,  Joule-Thomson coefficients, and the speed of sound are described in Section \ref{simulations}. In Section \ref{virial} we describe how the virial coefficients were determined and discuss the limits of the virial expansion. Section \ref{perturbation} is devoted to the Barker-Henderson perturbation theory, and Section \ref{EoS} includes a summary of the equations of state that we have considered in this work. Results for $PV$-isotherms, vapour pressure, the Joule-Thomson coefficient, and the speed of sound are presented in Section \ref{results} and discussed in light of the simulation results. Our evaluation of the EoS is also contained in this section.
Finally, conclusions are presented in Section \ref{conclusions}.

\section{MD and GEMC simulations}
\label{simulations}

\subsection{Equilibrium MD simulations of PV isotherms - fluid state}
\label{MDmethod}

A systematic set of thermodynamic states was generated to determine pressure-volume ($PV$) isotherms with equlibrium $NVT$ MD simulations. Eight isotherms were generated in a range from $T^*=k_\text{B}T/\varepsilon=2.0$ (supercritical) to $T^*=0.8$ (subcritical) with densities in the range from $n^*=N\sigma^3/V=0.001$ to $n^*=0.7$ where $k_{\text{B}}$ is Boltzmann's constant. The number of particles was $N=$ 27,648 for densities $n^* < 0.1$ and $N=$ 16,000 for densities $n^* \geq 0.1$. The MD cell aspect ratio was $L_x/L_y=L_x/L_z=4$. An additional five isotherms were generated in the critical region, $0.88 \leq T^* \leq 0.92$ and $0.2 \leq n^* \leq 0.5$ with $N=$ 128,000 particles.

In all cases, the system was divided into 16 layers of equal thickness, perpendicular to the $x$-direction. The two end layers and two layers in the middle of the MD box were thermostatted to the the same target temperature by a simple velocity rescaling algorithm \cite{ikeshoji1994}. (An illustration of the MD box is shown in Figure \ref{fig:solidliquid} for a two-phase non-equilibrium case.) The particles in the remaining layers followed the unperturbed equations of motion with the velocity Verlet algorithm. The MD box thus had a plane with mirror symmetry in its middle. In the postprocessing analysis, this symmetry was used to pool data from the left and right halves of the cell, giving 8 pooled layers, two of which were the heat baths. Of these 8 layers, the 4 central ones were assumed to be unaffected by the velocity rescaling algorithm and were used to collect data for the $PV$ isotherms.

Unless otherwise specified the time step was $\delta t^*=0.002$. The length of each run was one million time steps starting with a face-centred cubic (fcc) configuration. Subaverages over intervals of 50,000 were stored for postprocessing. The last 500,000 steps (10 dumps) were used for computing averages and uncertainties, which means that 4 layers times 10 dumps were used for computing averages.

\subsection{Equilibrium MD simulation of two-phase systems}
\label{eqmd}

Coexisting phases of gas and solid, and gas and liquid, were generated using a method à la Abraham and co-workers \cite{abraham1975}. For the solid/liquid coexistence, we used a method à la Hafskjold and Ikeshoji \cite{hafskjold1996}.

\subsubsection{System specifications and data acquisition}

The number of particles used in the two-phase simulation varied between 500 and 256,000, depending on the phases being simulated. Local thermodynamic properties were computed in layers or planes perpendicular to the $x$-axis. The number of layers or planes varied between 64 and 1,800. When layers were used, each one of them was treated as a control volume in which local values of thermodynamic properties were computed. The local kinetic temperature was computed from the equipartition theorem and the local density as a simple volume average of the number of particles. The local pressure was computed either with the usual virial equation \cite{frenkel2001} or with the layer-averaged method described by Ikeshoji \textit{et al.} \cite{ikeshoji2003}. For the two-phase simulations, we typically generated 2,000,000 time steps after equilibration for the averages.

As an alternative to computing local properties in layers with a certain volume, we computed the density profile using SPAM averaging as described by Hoover \cite{hoover2006}. In this method, the box is divided up in the $x$-direction by a number of equally spaced planes. The density at a given plane, labelled $\gamma$, was computed from
\begin{equation}
n(x=x_\gamma) =\frac{1}{L_y L_z} \sum_{i=1}^N mw (|x_i-x_\gamma |),
\end{equation}
where $w(x)$ is Lucy’s 1-dimensional weight function given by
\begin{equation}
w(x) =\frac{5}{4h} \left ( 1 + 3 \frac{x}{h} \right ) \left ( 1 - \frac{x}{h} \right )^3.
\end{equation}
In these simulations, we used a total of 1,800 planes and a smoothing length of $h^*=3$ (in LJ units). The two planes at either end of the MD box were treated as elastic walls while periodic boundaries operated in the $y$- and $z$-directions. Independent runs were conducted at different starting temperatures. The starting configuration was then relaxed using isokinetic MD (the Gaussian thermostat held the global temperature at its set point value). Following 1,000,000 equilibration steps, production runs of 4,000,000 steps were conducted. The co-existing densities were then obtained by averaging the high and low regions of the density profile. 

A complete set of the simulation conditions is given in the supplementary information.

\subsubsection{Gas/liquid and gas/solid simulations}
\label{gasliquid}

A box of liquid or solid was prepared using isokinetic molecular dynamics. After equilibration, the box was embedded in a larger simulation box, creating a ribbon of liquid or solid between empty layers. The aspect ratio of the resulting box was $L_x/L_y=L_x/L_z$ typically in the range from 3:1 to 6:1. Care had to be taken to avoid a rapid expansion of the liquid or solid when the mother box was embedded, which means that the process had to be started at a low temperature with essentially zero vapour pressure. For the gas/liquid case at $T^* > 0.7$ we did some simulations with $N$=256,000 in order to push the simulation as close as possible to the critical point. The initial configuration was in this case an fcc lattice at $T^*=0.45$. The temperature was gradually increased to $T^*=0.7$ and the system was allowed to equilibrate. Subsequent states were generated by each run starting from the final configuration from the previous run, again with a gradual temperature increase. During the temperature increase, the temperature in the less condensed phase (outer layers) were kept slightly higher than in the more condensed phase (central layers) to stabilise central phase. 

For the gas/solid simulations the density of the solid prior to the box expansion was adjusted to zero pressure in $x$-, $y$-, and $z$-directions in order to ensure a stress-free solid when the empty boxes were added. We did not succeed in doing $NPT$ equilibrium simulations by adjusting the three components of the pressure tensor in the solid to the the hydrostatic pressure in the liquid, so we ended up by adjusting the dimensions of the solid phase manually. The solid surface against the fluid was in all cases the [1,0,0] surface.

In most of the simulations, the system's center of mass was fixed to the center of the MD box by shifting the $x$-component of the coordinate system on a regular basis, while in some of the simulations, a holonomic constraint was employed to keep the center-of-mass fixed. This was achieved using Gauss’ principle of least constraint. 
The equations of motion are given by:
\begin{equation}
\frac{d \mathbf{r}_i}{d t} = \frac{\mathbf{p}_i}{m},
\end{equation}
\begin{equation}
\frac{d \mathbf{p}_i}{d t} = \mathbf{F}_i-\alpha_\text{p} \mathbf{p}_i-\lambda_\text{p} \mathbf{\hat{e}_x},
\label{eqn:dpdt}
\end{equation}
where $m$ is the particle mass, $\mathbf{r}_i$ is the particle position, $\mathbf{p}_i$ its momentum, $\mathbf{F}_i$ the total force acting on $i$, while $\mathbf{\hat{e}}_x$ is a unit vector directed along the positive $x$-axis. The Gaussian multipliers were evaluated instantaneously from the following formulae:
\begin{equation}
\alpha_\text{p}=\frac{\sum_{i=1}^N \mathbf{F}_i \cdot \mathbf{p}_i}{\sum_{i=1}^N \mathbf{p}_i \cdot \mathbf{p}_i},
\end{equation}
\begin{equation}
\lambda_\text{p}=\frac{1}{N} \left( \sum_{i=1}^N F_i^x - \alpha_\text{p} \sum_{i=1}^N p_i^x \right).
\end{equation}
In addition, a linear proportional feedback \cite{travis1995} was employed to prevent the global temperature from drifting due to numerical errors implicit in the integration of the equations of motion. This extra term took the form: $-\kappa (T-T_0 ) \mathbf{p}_i$, which was added to the right-hand-side of Eq. \eqref{eqn:dpdt}. The feedback constant, $\kappa$, was set equal to 10. Feedback terms were also required for the center-of-mass constraints. The following terms were added to the equations of motion for positions and momenta respectively: $-C \sum_{i=1}^N x_i \hat{\mathbf{e}}_x$ and $- D \sum_{i=1}^N \dot{x}_i \hat{\mathbf{e}}_x$, where $x_i$ is the $x$-coordinate of particle $i$. The dot above denotes the time derivative. The constants $C$ and $D$ were both set equal to 10.

\subsubsection{Liquid/solid}
\label{liquidsolid}

\begin{figure}[tbp]
\includegraphics[height=9cm, width=16cm]{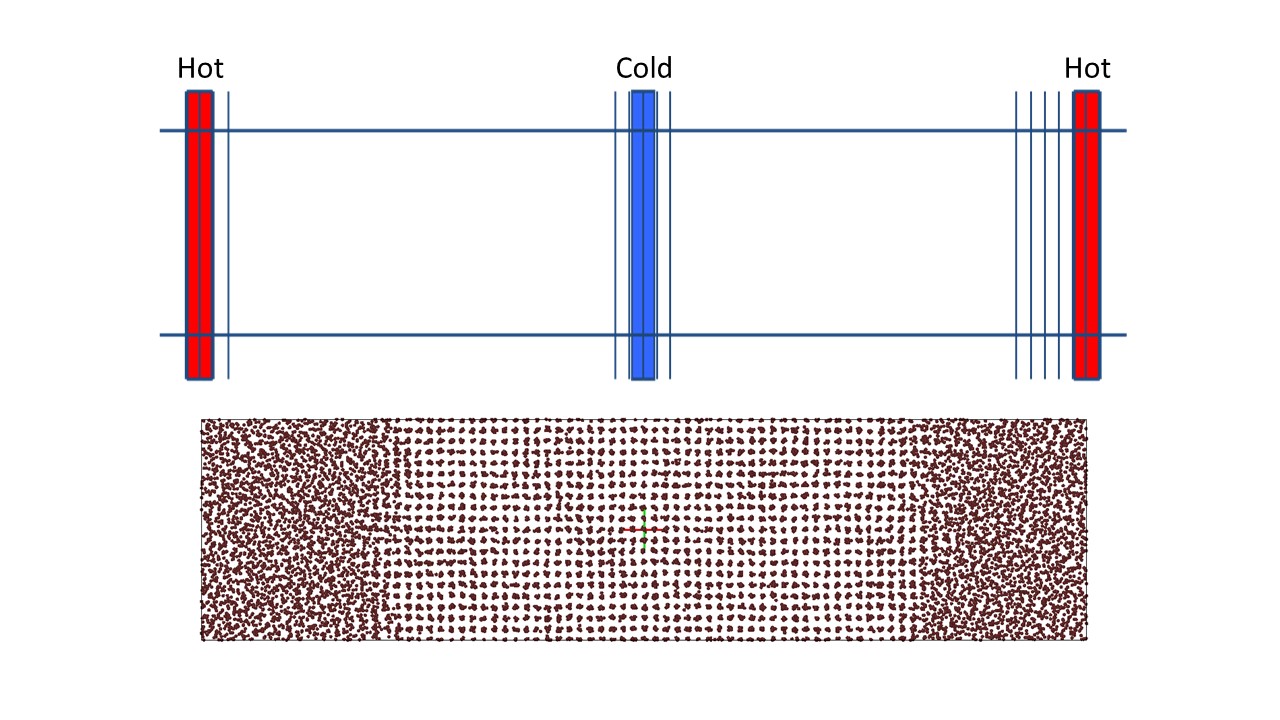}
\caption{Schematic view of the MD box with the thermostatted layers and a snapshot of a liquid-solid configuration.}
\label{fig:solidliquid}
\end{figure}

For the liquid/solid coexistence, we generated the system by starting with an equilibrium solid state and applied a temperature difference between the ends and the center of the MD box (in $x$-direction). The temperature difference was set up by thermostatting one layer at each end of the box to a high temperature and two layers in the center to a low temperature. The thermostatting was done by simple velocity rescaling with local momentum conservation \cite{ikeshoji1994}. If the overall density was within the two-phase region and the thermostat set points were chosen appropriately, the system would separate into two phases as shown in Figure \ref{fig:solidliquid}. After an initial period with different temperatures in the hot and cold regions, the thermostat set point at $T_\text{H}$ (hot) was changed to $T_\text{L}$ (cold). The system was then allowed to equilibrate with a $NVE$ simulation for a number of time steps. During the initial time steps with the temperature difference, the resulting temperature gradient was enough to stabilize the $x$-position of the liquid phase. During the subsequent equilibrium run, the center of mass of the whole system was fixed to the center of the MD box (in $x$-direction) by shifting the coordinate system. Also in this case the density had to be adjusted so that the solid phase in all three directions experienced the same stress in all three directions, given by the pressure in the fluid.

\subsection{Gibbs ensemble Monte Carlo}
\label{gemc}

The GEMC algorithm originally proposed by Panagiotopoulos \cite{panagiotopoulos1987} was used to obtain gas/liquid coexistence states. In this method, two simulation boxes undergo three types of moves: translations, isotropic expansion or contraction of the two boxes preserving the total system volume, and swapping a particle from one box to another.

A total of 1,500 particles were distributed over the two boxes. The configuration of the particles was sampled after each cycle, which consisted of, on average, 750 translation attempts, 750 swap attempts, and one volume move attempt. The step length for displacement and the volume moves were adjusted during equilibration to yield an acceptance ratio between 30\% and 50\%. The production run consisted of 20 blocks with 30,000 cycles each. 

\subsection{Simulations of the Joule-Thomson coefficient}

The Joule-Thomson (JT) coefficient describes the change in temperature experienced by a gas undergoing a throttling process at constant enthalpy. The JT coefficient is defined as
\begin{equation}
\mu_{\text{JT}}=\left( \frac{\partial T}{\partial P} \right)_H,
\label{eqn:JTdef}
\end{equation}
where $H$ and $V$ are the system's enthalpy and volume, respectively. It may be evaluated as
\begin{equation}
\mu_{\text{JT}}= \frac{V}{C_P} \left ( \alpha_P T - 1 \right),
\label{eqn:JTevalb}
\end{equation}
where $C_P$ is the heat capacity at constant pressure,
\begin{equation}
C_P=\left( \frac{\partial H}{\partial T} \right)_P,
\label{eqn:heatcapacity}
\end{equation}
and $\alpha_P$ is the thermal expansion coefficient at constant pressure,
\begin{equation}
\alpha_P=\frac{1}{V}\left( \frac{\partial V}{\partial T} \right)_P.
\label{eqn:thermalexpansion}
\end{equation}

\subsubsection{Using non-equilibrium MD (NEMD)}
\label{JTnoneq}

As all the quantities needed to compute the JT coefficient are mechanical properties, the coefficient is conveniently computed in MD simulations. The temperature derivatives in Eqs. \eqref{eqn:heatcapacity} and \eqref{eqn:thermalexpansion} are taken at constant pressure, and it would have been logical to carry out series of $NPT$ simulations at different temperatures and do the derivations numerically. Instead, we used two observations made in previous non-equilibrium MD simulations in which the system was subject to a gradient in composition \cite{hafskjold1996}, namely that the pressure quickly becomes constant over the system, and that the local values of the system's thermodynamic properties are consistent with the equilibrium equation of state. When the system is divided into a number of layers, each one of them may be considered a subsystem with properties equal to their bonafide equilibrium properties. This has the advantage that temperature derivatives at constant pressure can be computed for a range of states in one single run. The zero for $\mu_{\text{JT}}$ can then be determined very effectively. Hence, the Joule-Thomson coefficient was computed in the following way.

A system of $N$ LJ/s particles was simulated by imposing a thermal gradient on the system as described by Ikeshoji and Hafskjold \cite{ikeshoji1994}. We have previously shown that even with very strong gradients in the system, local equilibrium is achieved \cite{hafskjold1995criteria}.  In all the cases reported with this method, $N=27,648$. Each run was started from an fcc lattice configuration with 100,000 initial Monte Carlo steps to randomize the configuration, followed by 2,000,000 NEMD time steps. The last 1,000,000 steps were used for acquisition of stationary-state data.

\begin{figure}[tbp]
\includegraphics[height=9cm, width=16cm]{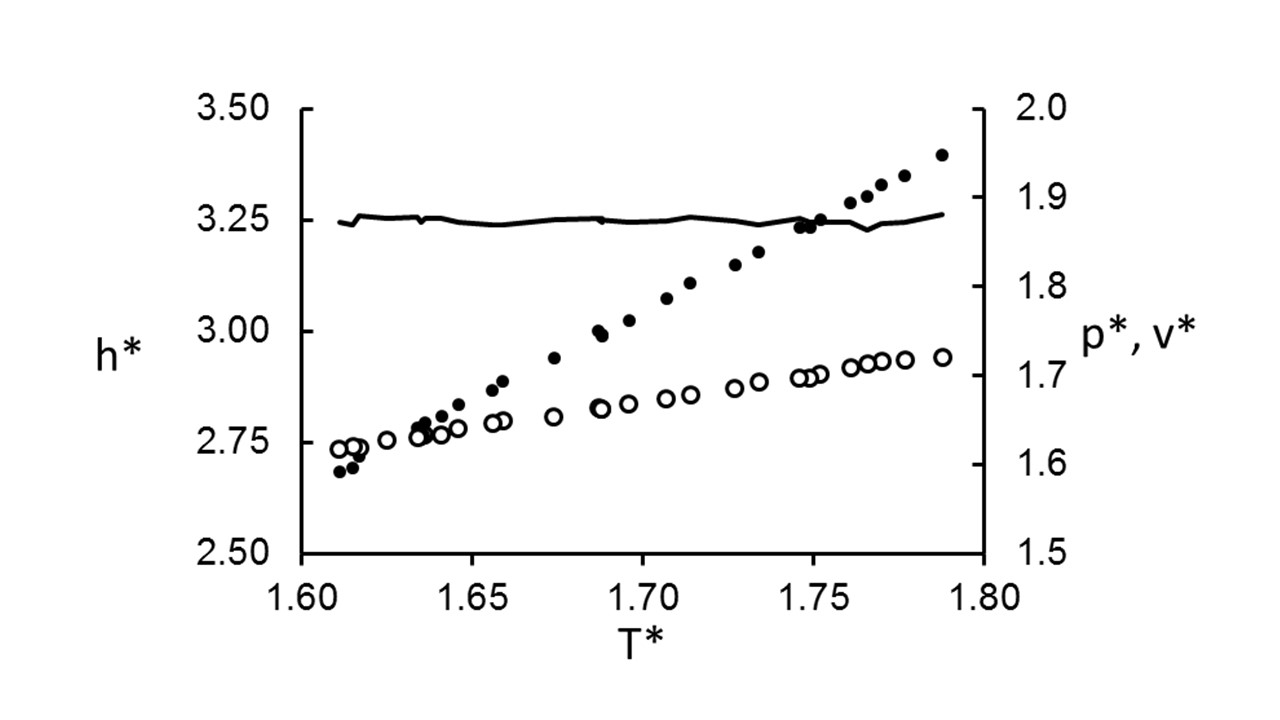}
\caption{Enthalpy (black dots) and volume (open symbols)  per particle as functions of temperature over half the MD box from a non-equilibrium simulation. The line shows that the pressure is constant along the box.}
\label{fig:thermalcoeffs}
\end{figure}

The temperature gradient, which typically spanned $\Delta T^*=0.2$ over half the MD box length, generated a density gradient at constant pressure in the system. An example of a non-equilibrium situation is shown in Figure \ref{fig:thermalcoeffs}.
Second-order polynomials were fitted
to the data, from which the temperature derivatives in Eqs. \eqref{eqn:heatcapacity} and \eqref{eqn:thermalexpansion} for the heat capacities and expansion coefficients were derived. The values of density and temperature corresponding to a given value
of the Joule-Thomson coefficient (\textit{e.g.} $\mu_{\textnormal{JT}}^{*}=0$)
were then determined from such plots by first fitting straight lines
to the data near the $\mu_{\textnormal{JT}}^{*}$-value of interest
and then determine the values for $T^{*}$ and $n^{*}$. Such plots were used to compute the Joule-Thomson coefficient from Eq. \eqref{eqn:JTevalb} for a range of densities and temperatures.
Contour lines for five different constant values of the Joule-Thomson coefficient  for $\mu^* \in \{-0.2,-0.1,0,0.1,0.2\}$ were computed with selected results for the JT inversion curve ($\mu^*=0$) shown in Section \ref{JTsimresults}.

\subsubsection{Using Isobaric-Isenthalpic Molecular Dynamics}
\label{isobaric}

From the definition of the inversion curve as the locus of points of vanishing JT coefficient, points on the curve can be obtained from the maxima of isenthalps. Using a refinement of the algorithm introduced by Kioupis and Maginn \cite{kioupis2002}, we used Nosé-Hoover dynamics to generate isenthalps from molecular dynamics.  The $6N+3$ equations of motion that generate the $NPH$ ensemble are:
\begin{subequations}
\begin{align}
\dot{\mathbf{r}_i} &= \frac{\mathbf{p}_i}{m} + \alpha_H \mathbf{r}_i, \\
\dot{\mathbf{p}_i} &=  \mathbf{F}_i-(\alpha_H + \gamma) \mathbf{p}_i, \\
\dot{V} &= 3V \alpha_H, \\
\dot{\alpha} &= 3V \beta_H [ P(t) - P_\text{req} ], \\
\dot{\gamma} &= \zeta [ H(t) - H_\text{req} ],
\end{align}
\end{subequations}
where $\beta_H$ and $\zeta$ are feedback constants which specify the strength of coupling of the system to the external reservoirs. Values of 5 and 10 (in LJ units) were used for $\beta_H$ and $\zeta$, respectively, throughout the simulations. To generate a particular isenthalp, a set of independent runs covering a fixed value of enthalpy but a range of pressures was conducted. For each simulation, an equilibration run of 100,000 time steps was first conducted, followed by a 2,000,000 step production run over which average values of thermodynamic properties were obtained. The system size was $N = 500$, the MD timestep was $\delta t^* =0.001$ and the equilibration runs were started from an fcc lattice.  The mean temperature was plotted as a function of pressure and then the states along the isenthalp were fitted to a polynomial. The stationary point was then obtained from the roots of the derivative of this polynomial. For the present work, a total of 25 isenthalps were generated, each one giving rise to a data point in the inversion curve for comparison with the MD method described above, perturbation theory, the virial expansion, and EoS.

\subsection{Speed of sound}
\label{speedofsound}

The speed of sound at zero frequency, $v_s$, in a fluid is given by \cite{hirschfelder1954}
\begin{equation}
v_s^2 = \left ( \frac{\partial P}{\partial n} \right )_S = \frac{C_P}{C_V} \left ( \frac{\partial P}{\partial n} \right )_T.
\label{eqn:soundspeed}
\end{equation}
This is a purely thermodynamic quantity which often appears in transport equations for fluids. 
The method used to compute the derivatives in Eq. \eqref{eqn:soundspeed} was essentially the same as the one described in Section \ref{JTnoneq}.
Three states were investigated, \textit{viz.} gas, supercritical fluid, and liquid. 

The quantity $\left ( \frac{\partial P}{\partial n} \right )_T$ (the inverse isothermal compressibility divided by the number density $n$) was computed as follows. For each state, a series of four equilibrium $NVT$-simulations were made at a given temperature and four different densities. In each case, $N=8,000$. Each run was started from an fcc-configuration and 1,000,000 Monte Carlo steps to equilibrate the system, followed by 4,000,000 MD steps where the last 2,000,000 steps were used to get the equilibrium data. Only the non-thermostatted layers in the MD box were used for acquisition of thermodynamic properties. The isothermal compressibility was then determined from a linear fit to the data for $P$ vs. $n$.

The heat capacity at constant pressure was computed as described in Section \ref{JTnoneq}.
The heat capacity at constant volume was determined in a similar way, except that the temperature was varied instead of the volume.

\section{The virial expansion}
\label{virial}

Mayers’ virial expansion for the pressure of a gas can be written as a power series in the density,
\begin{equation}
Z=\frac{\beta P}{n}=1+\sum_{k=2}^{\infty} B_k n^{(k-1)},
\label{eqn:virialexp}
\end{equation}
where the coefficients $B_k$ are multi-dimensional integrals involving products of Mayer $f$-functions, $f(r)=e^{-\beta u(r)}-1$. The coefficients depend on temperature, but are independent of density. In Eq. \eqref{eqn:virialexp}, $Z$ is the compression factor, $\beta$ is the inverse temperature from statistical mechanics, $\beta=\frac{1}{k_\text{B} T}$.

With the virial expansion for the compression factor, other thermodynamic properties may be derived through the Helmholtz free energy by  thermodynamic integration:
\begin{subequations}
\begin{align}
\beta A =&\ln (n\Lambda^3)-1+\int_0^n \frac{1}{n'} (Z - 1) dn' \\
=& \ln (n\Lambda^3) -1 +\sum_{k=2}^\infty \frac{1}{k-1}B_k n^{k-1},
\label{eqn:freeenergy}
\end{align}
\end{subequations}
where $\Lambda$ is the thermal de Broglie wave length. The corresponding expression for Gibbs energy per particle is
\begin{equation}
\beta G =\beta A + Z = \ln (\Lambda^{*3}) + \ln n^* +\sum_{k=2}^m \frac{k}{k-1} B_k n^{k-1},
\label{eqn:gibbs}
\end{equation}
where we have introduced the dimensionless quantities $\Lambda^* = \Lambda/\sigma$ and $n^* = n\sigma^3$. In the following, we shall discard the term $\ln (\Lambda^{*3})$ as this term plays no role in phase equilibria within the framework of classical statistical mechanics.

The lowest order coefficient, $B_2$, is given by the one-dimensional integral,
\begin{equation}
B_2=-2\pi \int_0^{r_c} f(r)r^2 dr,
\label{eqn:second}
\end{equation}
which can be performed analytically for a handful of pair potentials. This is not the case for the LJ/s model however, but a standard numerical quadrature is straightforward for this short ranged potential.

The higher order coefficients become progressively harder to represent mathematically and evaluate numerically. The multidimensional integrals are conveniently represented as cluster diagrams using graph theoretic concepts. In this format, $B_3$ can be represented by a single Mayer (cluster) diagram, $B_4$ requires 3 while $B_5$ involves a sum of 10 diagrams. These can be reduced in number by switching from Mayer diagrams to so–called Ree-Hoover diagrams \cite{ree1964}.

The second virial coefficient, $B_2$, was calculated using 16-point Gauss-Legendre quadrature, while higher coefficients were calculated using a hit-and-miss Monte Carlo scheme. 
Briefly, the method for estimating $B_3$ was as follows: a particle was placed at the origin and a second particle was placed randomly within a sphere of radius $R$ ($=r_c$ in this work), centred on the first particle. 
The third particle was likewise placed randomly within a sphere of radius $R$, also centred on particle 1. With the three particles so placed, the Mayer $f$-functions were calculated and then sums of the products $f_{12}f_{13}f_{23}$ and $f_{12}f_{13}$ were compiled over 10 billion Monte Carlo trials. The third virial coefficient was then obtained as
\begin{equation}
B_3 = -\frac{4}{3} \frac{\langle f_{12}f_{13}f_{23}\rangle}{\langle f_{12}f_{13} \rangle }.
\end{equation}
This algorithm was extended in an obvious way to compute $B_4$ and validated against literature results for hard spheres.

\section{Perturbation theory}
\label{perturbation}

We have used Barker-Henderson Perturbation theory \cite{barker1967I,barker1967II} to obtain thermodynamic properties of the LJ/s model. In this theory, the potential is split at the point where the potential is zero and the free energy is expanded in powers of a depth parameter \cite{barker1967II}. The reference system is approximated by a hard-sphere fluid with an effective (temperature dependent) diameter given by
\begin{equation}
d= \int_0^\sigma \left ( 1 - e^{-\beta \phi (r) } \right ) dr.
\end{equation}

The hard-sphere diameter determined in this way for the LJ/s model is exactly the same as for the LJ model.
The full residual Helmholtz free energy, correct to second order in the perturbation, is then given by
\begin{equation}
A = A_0 + \beta A_1 + \beta^2 A_2,
\label{eqn:pt}
\end{equation}
where $A_0$ is the free energy per particle of the hard-sphere reference system. $A_1$ is the first-order perturbation term, representing the mean attractive energy, and $A_2$ is the second-order perturbation term representing the fluctuation contribution. The first-order term is calculated from
\begin{equation}
A_1 = 2 \pi n k_\text{B} T \int_\sigma^\infty g_0 (r) u(r) r^2 dr.
\label{eqn:pt_a1}
\end{equation}
Here $g_0 (r)$ is the radial distribution function of the hard-sphere reference system. 

The second-order perturbation term is under the so-called improved macroscopic compressibility approximation expressed as \cite{zhang1999}
\begin{equation}
A_2 = -\pi K_T^0 (1+\chi) n k_\text{B} T \int_\sigma^\infty g_0 (r) u^2 (r) r^2 dr.
\label{eqn:pt_a2}
\end{equation}
The factor $K_T^0=k_\text{B}T\left(\frac{\partial n}{\partial P} \right )_T$ is the isothermal compressibility of the hard-sphere system reduced by the ideal gas isothermal compressibility, $\left(k_\text{B} T n\right)^{-1}$,  and $\chi$ is a correction factor justified by the argument that molecules in neighbouring coordination shells are
correlated. In this work we apply macroscopic compressibility approximation (MCA), setting $\chi =0$ \cite{hansen2013}. We shall refer to a theory that includes only the first two terms on the right-hand side of Eq.~\eqref{eqn:pt} as a first-order perturbation theory, and a theory that includes all three terms and the MCA as a second-order perturbation theory.

The isothermal compressibility is obtained from the Carnahan-Starling (CS) equation of state \cite{carnahan1969},
\begin{equation}
Z_0^{\text{CS}} = \frac{PV}{Nk_\text{B}T} = \frac{1 + \xi+\xi^2-\xi^3}{(1-\xi)^3},
\label{eqn:cs}
\end{equation}
where $\xi= \pi n d^3/6$ is the hard-sphere packing fraction:
\begin{equation}
K_T^{0,\text{CS}} = \frac{(1-\xi)^4}{1+4\xi+4\xi^2-4\xi^3+\xi^4}.
\label{eqn:cscompressibility}
\end{equation}
Applied to the CS equation of state, we arrive at the excess free energy for the reference system \cite{heyes1998}
\begin{equation}
A_0^{\text{CS}} = k_\text{B} T \frac{(4 - 3 \xi)\xi}{(1-\xi)^2},
\end{equation}
where we have again omitted terms that are independent of the density or packing fraction. With the free energy of the reference system and the perturbation expansion, other thermodynamic functions may be derived from the Helmholtz free energy. 

In the implementation of Eqs.~\eqref{eqn:pt_a1} and \eqref{eqn:pt_a2}, we computed the integrals on the right-hand side numerically, where the hard-sphere radial distribution function was computed as described in Ref.~\cite{Lafitte2013}. A reduced form of the integral part of $A_1$ and $A_2$ is correlated in density and the temperature-dependent ratio $x_0=d/\sigma$. The reduced integrals, $a_1^s$ and $a_2^s$, are implicitly defined from,
\begin{align}
A_1 &= 2 \pi n^* k_\text{B} T \epsilon a_1^s, \label{eqn:pt_a1s} \\
A_2 &= -\pi K_T^0 (1+\chi) n^* k_\text{B} T \epsilon^2 a_2^s,\label{eqn:pt_a2s}
\end{align}
and are correlated using the functional form,
\begin{align}
a_i^s = &p_{i,1}\left(n^*\right)^4+p_{i,2}\left(n^*\right)^3+p_{i,3}\left(n^*\right)^2+p_{i,4}n^*+p_{i,5} \nonumber\\
&+n^*\left(p_{i,6}\left(n^*\right)^2+p_{i,7}n^*+p_{i,8}\right)\left(x_0-1\right) \nonumber\\
&+n^*\left(p_{i,9}\left(n^*\right)^2+p_{i,10}n^*+p_{i,11}\right)\left(x_0-1\right)^2.
\label{eq:a1spoly}
\end{align}
The two-dimensional polynomials for $a_1^s$ and $a_2^s$  were regressed against 5 isotherms in the domain of interest for this work, $T^* \in \{0.4,0.7,0.85,1.0,2.0\}$ and $0<n^*<0.9$, such that the correlations reproduced results from the numerical integration within a sufficiently high accuracy that the two formulations could not be distinguished from each other when plotted as a function of the reduced density. 

The obtained parameters for Eq. \eqref{eq:a1spoly} are given in Table \ref{tab:par1}.
\begin{table}
  \centering
  \caption{The $a_1^s$ and $a_2^s$ parameters}
  \begin{tabular}{ c  c  c }
    \toprule
    $j$ of $p_{i,j}$	&	$a_1^s$	&	$a_2^s$	\\
    \toprule
    1	&	0.04605	&	-0.1124		\\
    2	&	0.4554	&	-0.2830		\\
    3	&	-0.3328	&	0.3318		\\
    4	&	-0.3464	&	0.2507		\\
    5	&	-0.5351	&	0.3585		\\
    6	&	-7.529	&	6.794		\\
    7	&	9.489	&	-9.266		\\
    8	&	0.5337	&	0.1336	\\
    9	&	30.73	&	-27.37		\\
    10	&	-45.88	&	42.43		\\
    11	&	4.627	&	-5.655 \\
    \bottomrule
  \end{tabular}
  \label{tab:par1}
\end{table}
The analytic representation of the first and second order perturbation theories as well as the analytic derivatives up to second order were next implemented in the thermodynamic framework described in Ref.~\cite{Wilhelmsen2017}, which allows thermodynamic properties to be computed with a high accuracy, and phase-equilibrium calculations to be performed with state-of-the-art algorithms~\cite{Aasen2017}.

\section{Cubic equations of State (EoS)}
\label{EoS}

The most commonly used cubic EoS are the Peng-Robinson~(PR)~\cite{Peng_1976} and Soave-Redlich-Kwong~(SRK)~\cite{SRK}. They are frequently used to describe the properties of fluids, \textit{e.g.} in process simulations, computational fluid dynamics or demanding optimisation studies. The generic cubic EoS can be represented as:
\begin{equation}
P=\frac{R_gT}{v-b}-\frac{a\alpha_c(T)}{\left(v-bm_1\right)\left(v-bm_2\right)},
\label{eqn:PRpressure}
\end{equation}
\noindent 
where $P$ is the pressure, $R_g$ the universal gas constant, $v$ the molar volume, and $a$, $\alpha_c$, and $b$ are parameters of the EoS. The constants $m_1$ and $m_2$ characterize various two-parameter cubic EoS. For PR, $m_1=-1+\sqrt{2}$ and $m_2=-1-\sqrt{2}$, while for SKR $m_1=-1$ and $m_2=0$. Eq. \eqref{eqn:PRpressure} can be integrated with respect to the volume to give the excess Helmholtz energy density. The Helmholtz energy can be further differentiated to give thermodynamic variables. The input parameters to SRK and PR are the critical temperature, pressure, and the acentric factor, defined as:
\begin{equation}
\Omega=-\log_{10}\left(P_{\text{r},0.7}\right) - 1,
\label{eq:acf}
\end{equation}
where $P_{\text{r},0.7}$ is the reduced pressure $P_\text{r}=P/P_\text{critical}$ at a reduced temperature of $T_\text{r}=T/T_\text{critical}=0.7$. While $a$ and $b$ can be determined solely by the critical temperature and pressure of the fluid, the parameter $\alpha_c$ is a function of the reduced temperature:
\begin{equation}
    \alpha_c=\left(1+\kappa_c\left(1-\sqrt{T_\text{r}}\right)\right)^2.
\end{equation}
Here $\kappa_c$ is a second order polynomial in the acentric factor with coefficients from Graboski and Daubert~\cite{Graboski1978} for SRK and the standard parameters for PR~\cite{Peng_1976}. Hence, it is necessary to determine the critical temperature, pressure, and the acentric factor by use of simulations in order to use cubic EoS to compute the thermodynamic properties of the LJ/s model.

\section{Results}
\label{results}

\subsection{$PV$ isotherms and compression factor}

\begin{figure}[tbp]
\includegraphics[height=11cm, width=15cm]{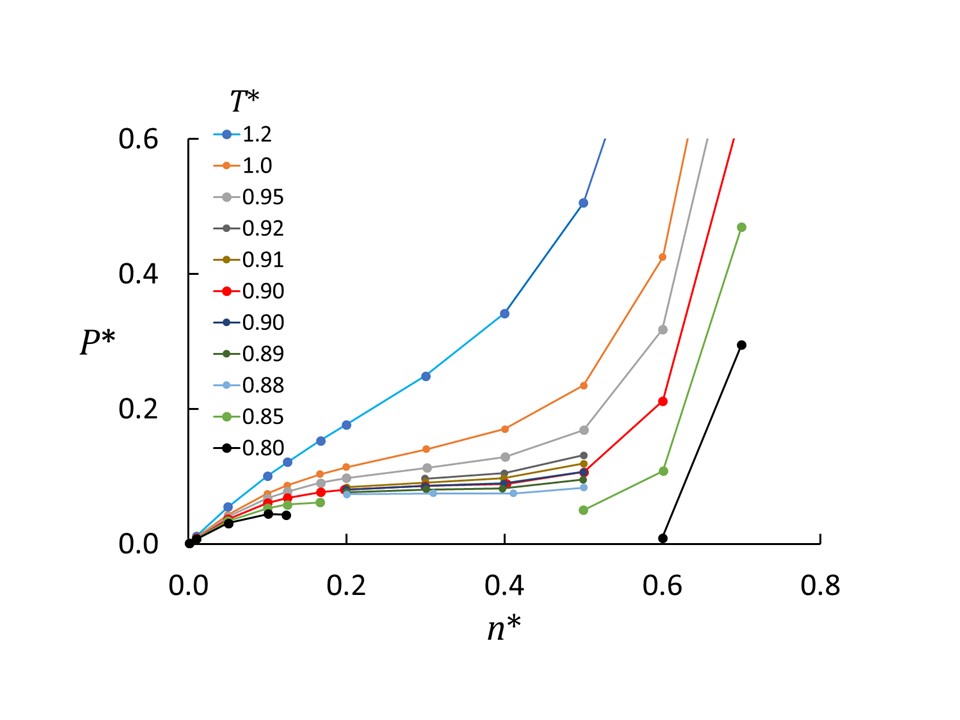}
\caption{Selected isotherms from MD simulations. The pressure is almost constant for $0.88 < T^* < 0.90$ for $0.2 < n^* < 0.4$, indicating that the gas/liquid critical point is in this region. Broken lines for $T^* \leq 0.85$ indicate that data are omitted in the two-phase region. Three standard errors, determined as described in Section \ref{MDmethod}, are smaller than the symbol size.}
\label{fig:Pnisotherms}
\end{figure}

\subsubsection{Simulations}
Figure \ref{fig:Pnisotherms} shows selected isotherms in the fluid regime resulting from the MD simulations. Isotherms at $T^* > 0.9$ are clearly supercritical and those at $T^* = 0.85$ and $0.80$ are subcritical. Some of the densities for the subcritical isotherms gave phase separation and are omitted in the figure. The isotherms at $T^* \approx 0.88 - 0.89$ are essentially horizontal around $n^* \approx 0.3$, indicating that the critical point lies in this region. Visualisation of the final configurations shows that the isotherm at $T^*= 0.90$ is very close to critical, whereas those at $T^* < 0.85$ are subcritical. 

Numerical values for the pressure, internal energy, enthalpy, and compression factor are given in the supplementary information. The table includes the computed temperature as described in Section \ref{MDmethod} and the other properties as defined in the table caption.

\subsubsection{The virial expansion}
\label{resultsvirial}

The virial coefficients $B_2$, $B_3$, and $B_4$ were computed as described in Section \ref{virial} for temperatures in the range $0.2 < T* < 4.0$. The parameters $a_{k,n}$ in the model
\begin{equation}
B_k=\sum_{n=0}^m a_{k,n} T^{-n}
\label{eqn:virialcoeff}
\end{equation}
were fitted to the data with the result shown in Table \ref{virial_fit_coeff}.
\begin{table}
\centering
\caption{Fitted coefficients $a_{k,n}$ of the inverse temperature relation, Eq. \eqref{eqn:virialcoeff} for the virial coefficients $B_2,B_3$, and $B_4$. The uncertainties represent 95\% confidence intervals.}
\vspace{0.4 cm}
\begin{tabular}{c c c c}\hline
$n$ & $k=2$ & $k=3$ & $k=4$ \\
\hline
$0$ & $1.345\pm 0.007$  & $3.76 \pm 0.02$  & $-1.376 \pm 0.007$ \\
$1$ & $-1.336 \pm 0.007$  & $  -20.9 \pm 0.1 $ & $ 40.5 \pm 0.2 $ \\
$2$ & $-3.85 \pm 0.02$ & $ 64.0 \pm 0.3 $ & $ -260 \pm 1 $ \\
$3$ & $1.295 \pm 0.006$ &  $ -90.2 \pm 0.5 $ & $ 936 \pm 5 $ \\
$4$ & $-0.416 \pm 0.002$ &  $ 66.8 \pm 0.3 $ & $ -2069 \pm 10 $  \\
$5$ & $-$ & $ -20.1 \pm 0.1 $ & $ 2789 \pm 14 $ \\
$6$ & $-$ & $-$ & $ -2240 \pm 11 $ \\
$7$ & $-$ & $-$ & $ 1010 \pm 5 $ \\
$8$ & $-$ & $-$ & $ -200 \pm 1 $ \\\hline
\end{tabular}
\label{virial_fit_coeff}    
\end{table}

\begin{figure}[tbp]
\includegraphics[height=11cm, width=15cm]{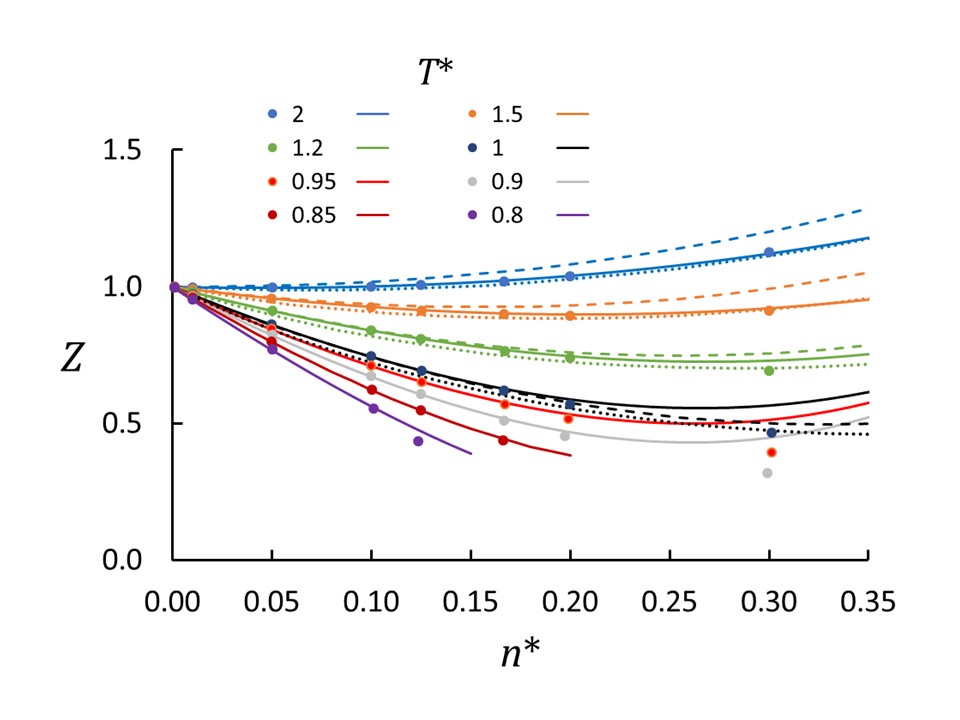}
\caption{Compression factor isotherms from the virial expansion including $B_2 - B_4$ (lines) compared with results from MD simulations (dots). The two isotherms $T^*=0.8, 0.85$ are subcritical.  Three standard errors in the MD data, determined as described in Section \ref{MDmethod}, are smaller than the symbol size. Included in the figure are also results from the SRK (dashed lines) and PR (dotted lines) EoS for the four highest temperatures.}
\label{fig:Z}
\end{figure}

The compression factor is compared with MD results for the low-density region in Figure \ref{fig:Z}. The agreement is excellent for the high temperatures shown, but worsens close to the critical temperature. Using Eq. \eqref{eqn:gibbs} with virial coefficients $B_2$ to $B_4$, we estimate the critical temperature to $T^* \approx 0.83$, the critical pressure to $P^* \approx 0.08$, and the critical density to $n^* \approx 0.19$, which is in fair agreement with the simulation results (see Section \ref{binodal}) considering the fact that the virial expansion is rather poor in the critical region (\textit{c.f.} Figure \ref{fig:Z}).

\subsubsection{Cubic equations of state}

Figure \ref{fig:Z} shows that the SRK and PR are accurate for low and intermediate densities with PR being slightly superior for the higher temperatures.
At low densities, both equations are inferior to the virial expansion, with SRK slightly better than PR. The inaccuracies in the low-density limit have the effect that the Joule-Thomson inversion curve is not well represented by SRK and PR in this limit, see Section \ref{resultsJTvirial}. The JT coefficient is sensitive to the temperature dependency of the second virial coefficient in this limit, which is not sufficiently accurate. For $n^* > 0.2$ the PR EoS becomes the more accurate of the two.

\subsection{The phase diagram}
\label{binodal}

The gas/liquid binodal was obtained from the requirement that for two phases to be in thermodynamic equilibrium, they must have the same temperature (thermal equilibrium), the same pressure (mechanical equilibrium) and the chemical potential of every component must be identical in the phases (thermodynamic equilibrium). For a one-component system, equality of chemical potentials is equivalent to equality of the Gibbs free energy per particle, $G$, of each phase. Thus, along an isotherm, the corresponding points on the gas/liquid binodal are obtained by solving the set of equations:
\begin{subequations}
\begin{align}
T_\text{g}&=T_\text{l} \text{    (isotherm)} \\
P_\text{g}&=P_\text{l} \\
G_\text{g} &= G_\text{l}
\end{align}
\end{subequations}
where the subscripts "g" and "l" respectively refer to the gas and liquid phases.

\subsubsection{Simulation results}
\label{simres}

\begin{figure}[tbp]
\includegraphics[height=11cm, width=15cm]{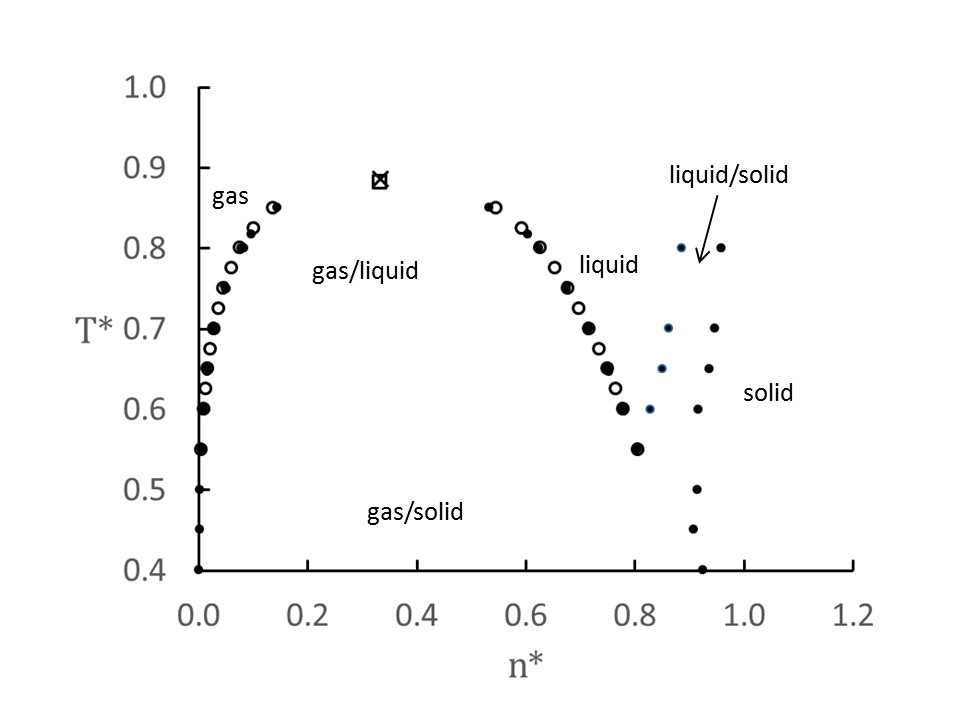}
\caption{Gas/liquid, liquid/solid, and gas/solid coexistence curves obtained with direct MD two-phase simulations (LAMMPS and NEMD, filled circles) and GEMC (open circles). The gas/liquid critical point was determined from MD (square) and GEMC (cross). The statistical uncertainties are smaller than the symbol size.}
\label{fig:phasediagram}
\end{figure}

The gas/liquid, gas/solid, and liquid/solid coexistence curves were determined as described in Sections \ref{eqmd} and \ref{gemc} with the results shown in Figure \ref{fig:phasediagram}. The agreement between MD and GEMC is excellent for the gas and liquid branches of the binodal curve. The results from LAMMPS \cite{plimpton1995} and NEMD\footnote{NEMD is used both as an acronym for "non-equilibrium molecular dynamics" and as the name of an in-house code.} are  indistinguishible for the binodal curve (LAMMPS was not used for the liquid/solid equilibrium).

\begin{figure}[tbp]
\includegraphics[height=11cm, width=15cm]{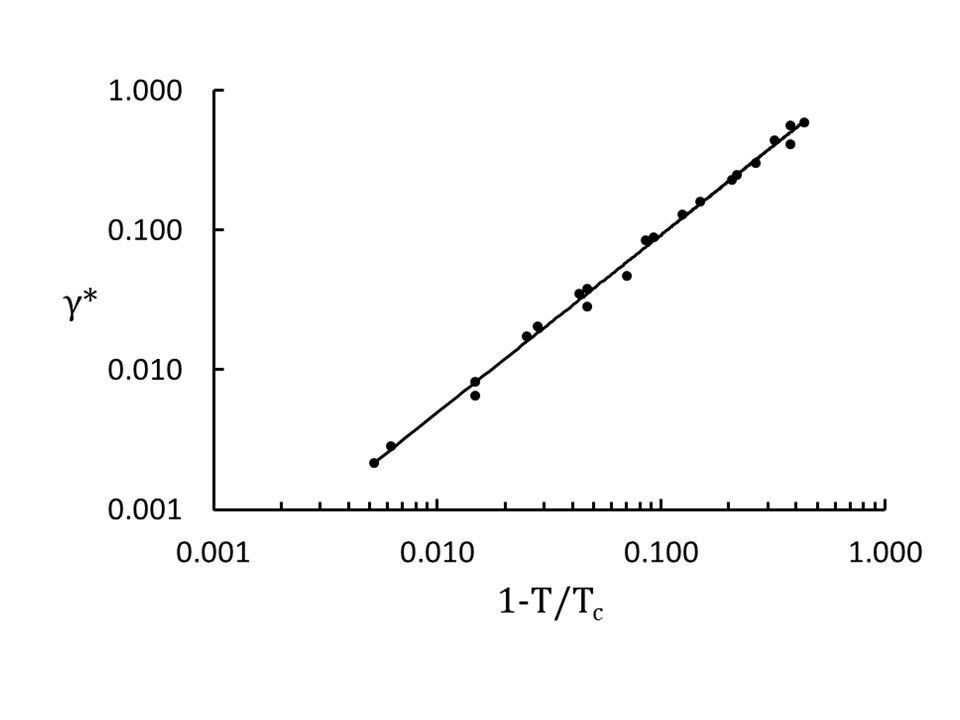}
\caption{The gas/liquid surface tension as function of the deviation from critical temperature.}
\label{fig:surfacetension}
\end{figure}

The critical temperature was determined from the MD data with a scaling-law analysis of the surface tension, $\gamma$, as function of $\epsilon = 1-T/T_\text{c}$, where $T_\text{c}$ is the estimated critical temperature. The behaviour is expected to be of the form $\gamma \sim \epsilon^{2\nu}$ where $\nu$ is a critical exponent with value $\nu \approx 0.64$ for a 3D system \cite{stanley1971}. The graph in Figure \ref{fig:surfacetension} shows MD data and a line of the form $\gamma ^* = A(1-T/T_\text{c})^B$ with $A$, $B$, and $T_\text{c}$ fitted to the MD data. The fitted critical temperature is $T_\text{c}^*=0.882$ and the fitted exponent is $B=1.27$ The latter value is in excellent agreement with the theoretical value $2\nu=1.26$ for the Ising universality class. With the critical temperature so determined, the critical density was determined from the rectilinear diameter to $n_\text{c}^* = 0.332$. The critical pressure was estimated by extrapolating a least-square fit of the Antoine equation $\ln P = A + B/T$ to the critical temperature with the result $P_\text{c}^*=0.074$.

It has already been well established that the LJ system belongs to the Ising universality class (see \textit{e.g.} Watanabe \textit{et al.} \cite{watanabe2012} and references therein). Our result for the value of the exponent $\nu$, suggests that this is also the case for the LJ/s system.

The gas/liquid critical point was also determined from the GEMC data in a similar way, except that we used the densities of the coexisting phases instead of the surface tension.
Following previous work by Vega \textit{et al.} \cite{Vega1992}, the critical temperature was estimated using the renormalization group scaling law for coexisting densities, $\rho_l - \rho_g \sim \epsilon^\beta$. Following Vega \textit{et al.}, a least-squares fit to the GEMC data in the range $\epsilon < 0.2$ and with a fixed $\beta=0.325$ gave $T_c=0.887$. Based on the GEMC data, the critical pressure was estimated as described above to $P_c^*=0.075$. The critical density was found to be $n^*=0.333 \pm 0.001$, again from the rectilinear diameter.

By comparison, the critical temperature of the LJ model is in the range 1.311 - 1.313 and the critical density is 0.316 - 0.317 \cite{perez2006}.
The difference between the critical temperature of the LJ and LJ/s models is clearly due to the more long-range attraction in the LJ fluid leading to more cohesive energy. The difference in critical density is small and the higher critical density of the LJ/s fluid may follow from the lower critical temperature.

\begin{figure}
\includegraphics[scale=0.5]{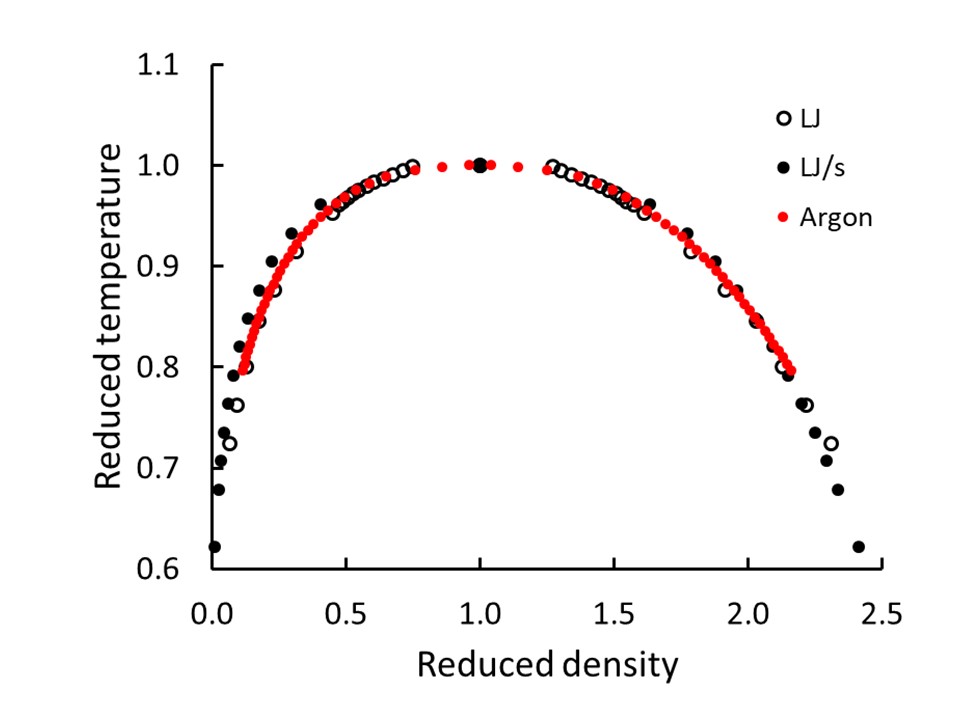}\caption{Binodal curves for LJ, LJ/s, and argon in units scaled by the critical temperatures and densities.}
\label{fig:binodalreduced}
\end{figure}

The comparison with the LJ binodal curve gives a different impression when plotted in reduced units, \textit{i.e.} scaled with the critical temperature and density. This is shown in Figure  \ref{fig:binodalreduced}, which also includes experimental data for argon \cite{michels1958}. The data for the LJ binodal are from Potoff and Panagiotopoulos \cite{potoff2000}. Data from Watanabe \textit{et al.} \cite{watanabe2012} for a smoothly truncated LJ potential at $r_\text{c}=3.0$ are virtually on top of the argon data (not shown). We conclude that the LJ/s and other truncated LJ potentials do an equally good job in modelling argon in the critical region as does the full LJ model. This confirms the picture drawn by Orea \textit{et al.} \cite{orea2015} that the law of corresponding states is a very strong scaling law and should be a useful tool in understanding the role of the attractive tail in pair potentials.

The triple-point temperature for LJ/s was determined from the data shown in Figure \ref{fig:phasediagram}. Following Ahmed and Sadus \cite{ahmed2009}, we determined the triple point temperature as the intersection of the liquid branch of the binodal curve and the liquidus line with the result $T_\text{tp}=0.547 \pm 0.005$. This is lower that the value for the LJ model, which has been found to be in the range 0.66 - 0.70 \cite{ahmed2009}. For the LJ/s model, we found the coexisting liquid and solid densities at the triple point to be $0.80 \pm 0.01$ and $0.90 \pm 0.02$, respectively. This is also lower than for the LJ model, which are 0.86 and 0.98, respectively.

Numerical values for the coexistence curves are given as supplementary information.

\subsubsection{Perturbation theory}
\label{pt}

\begin{figure}
\includegraphics[scale=0.5]{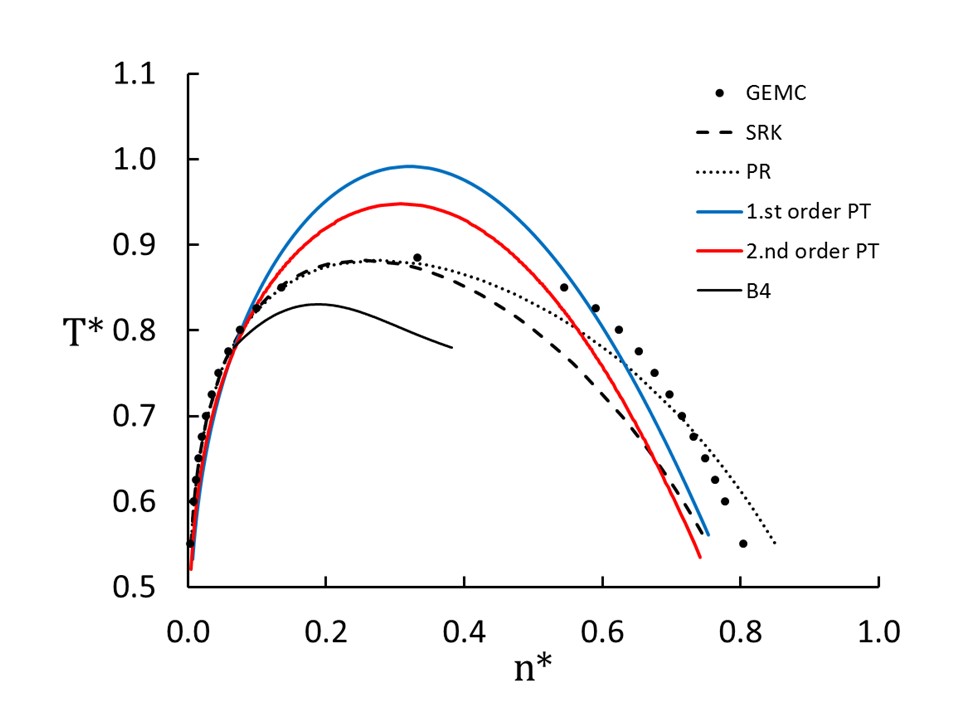}\caption{Binodal curves as determined by SRK, PR, virial expansion, and $1^\text{st}$ and $2^\text{nd}$ perturbation theory compared with the GEMC simulation data.}
\label{fig:cubicsatdens}
\end{figure}

Figure~\ref{fig:cubicsatdens} shows predictions of the first- and second-order perturbation theories where the MCA was invoked for the second-order theory. Somewhat surprisingly, the second-order perturbation theory predicts densities of the liquid-phase that are further away from the simulation data than the first-order theory. The critical temperature, however, is closer to the simulations for the second-order theory. This can be explained on the basis of the MCA that was invoked in the second-order perturbation theory. The second-order perturbation term of the LJ fluid is compared to MC results in Fig 2B in the work by van Westen and Gross~\cite{van2017}, where it is shown that the MCA becomes increasingly worse at higher densities. Hence, for the LJ/s model, it is apparently better to rely on the first-order theory as the MCA seems to move the coexistence liquid-phase densities away from the simulation data. 

Figure 10 in the work by van Westen and Gross \cite{van2017} compares the agreement between the perturbation theories and simulations with increasing order of the perturbation theory. Already the first-order theory predicts a critical temperature which is around 1.38, where the simulations extrapolate to 1.31, \textit{i.e.} 5\% higher than simulations. The agreement is much worse for the LJ/s model, where the first-order theory gives a critical temperature of 0.99, \textit{i.e.} about 12\% higher than that from simulations.

It is known from studies of the square-well potential that a shorter range of the potential makes the perturbation theory less accurate \cite{Vega1992}.
It is therefore to be expected that the perturbation theory is less accurate for the LJ/s model than for the LJ model. 

\subsubsection{Cubic equations of state}

The critical temperature and pressure of the LJ/s model were reported in Section \ref{simres}. From the vapour pressures reported in Section \ref{resultsPvap}, we estimate the acentric factor (see Eq.~\eqref{eq:acf}) to be $0.07 \pm 0.02$. This value is quite high for a system of spherical particles and different different from the LJ value, which is $\approx -0.036$ based on data from Johnson \textit{et al.} \cite{Johnson_1993}. 

The critical data and acentric factor enable us to evaluate the thermodynamic properties of the LJ/s model with cubic EoS.
Figure~\ref{fig:cubicsatdens} shows that both cubic EoS predict well the saturation density of the vapour phase, but are rather poor for the density of the liquid phase. This is a well-documented behaviour for cubic EoS, which can also be observed for other substances as well, see \textit{e.g.} Ref. \cite{Aasen2017}. Fig.~\ref{fig:cubicsatdens} shows that both SRK and PR under-predict the liquid-phase density at $T^*>0.7$, but the PR EoS less so than the SRK. Whereas the SRK liquid density is too low at all temperatures, the PR crosses over the simulation results and over-predicts the liquid-phase density for $T^*<0.7$. On this basis, one may argue that the PR is slightly better that the SRK, but none of them give very accurate predictions of the saturation densities.

\subsection{The vapour pressure}
\label{resultsPvap}

\begin{figure}
\includegraphics[scale=0.6]{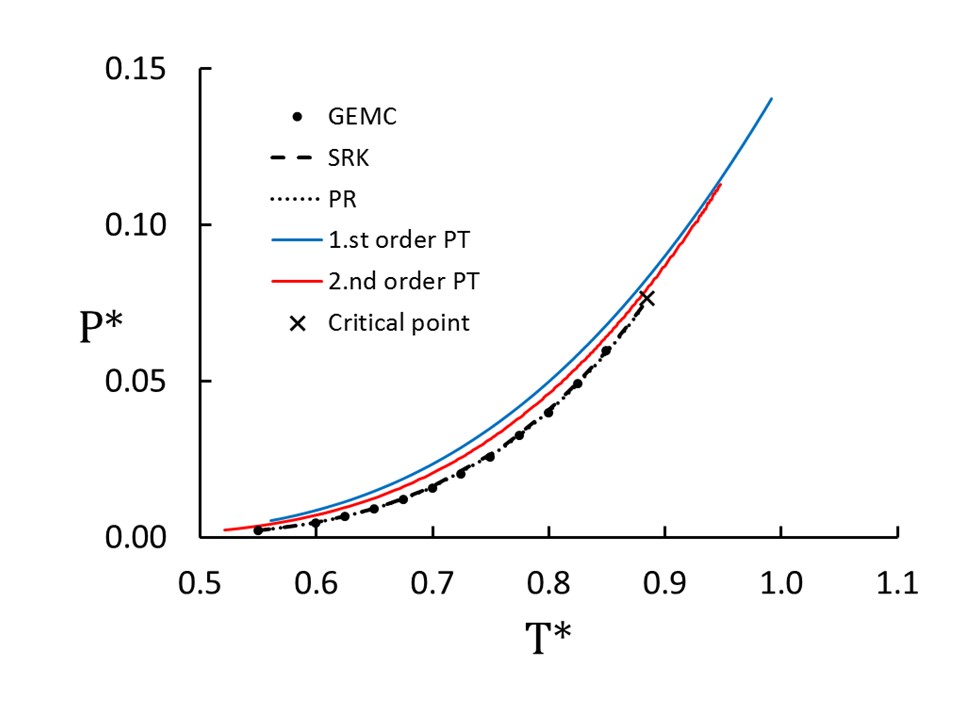}
\caption{Vapour pressure from SRK, PR, and $1^\text{st}$ and $2^\text{nd}$ order BH perturbation theory compared with the GEMC results. The critical point was determined by GEMC. Note that the cubic EoS parameters are determined such that the vapour pressure matches the GEMC data at $P_\text{r}=P/P_\text{critical}=1.0$ and $P_\text{r}=0.7$.}
\label{fig:cubicsatpres}
\end{figure}

The vapour pressure is shown as function of temperature in Figure \ref{fig:cubicsatpres}, where the results from the two cubic EoS and the first- and second-order perturbation theories are compared with GEMC data. Unlike the case for the liquid densities, we find that the second-order theory does a slightly better job than the first-order theory in matching the saturation pressures. Both cubic EoS reproduce the saturation pressures of the LJ/s model nearly within the accuracy of the simulations. This is a consequence of the way the EoS parameters are parameterized, since they reproduce by construction the critical temperature, pressure and saturation pressure at a reduced temperature of 0.7.

\subsection{The Joule-Thomson coefficient}
\label{jtcoeff}

The locus of points of vanishing JT coefficients defines the inversion curve, which can be determined as
\begin{equation}
T\left(\frac{\partial P}{\partial T}\right)_n-n\left(\frac{\partial P}{\partial n}\right)_T=0
\label{eqn:JTeval}
\end{equation}
This is a special case of $\mu_{\text{JT}} = constant$, which represents loci in the $\{T,P\}$ or $\{T,n\}$ plane with a fixed value of $\mu_{\text{JT}}$.

\subsubsection{Simulations}
\label{JTsimresults}

\begin{figure}
\begin{raggedright}
\includegraphics[scale=0.8]{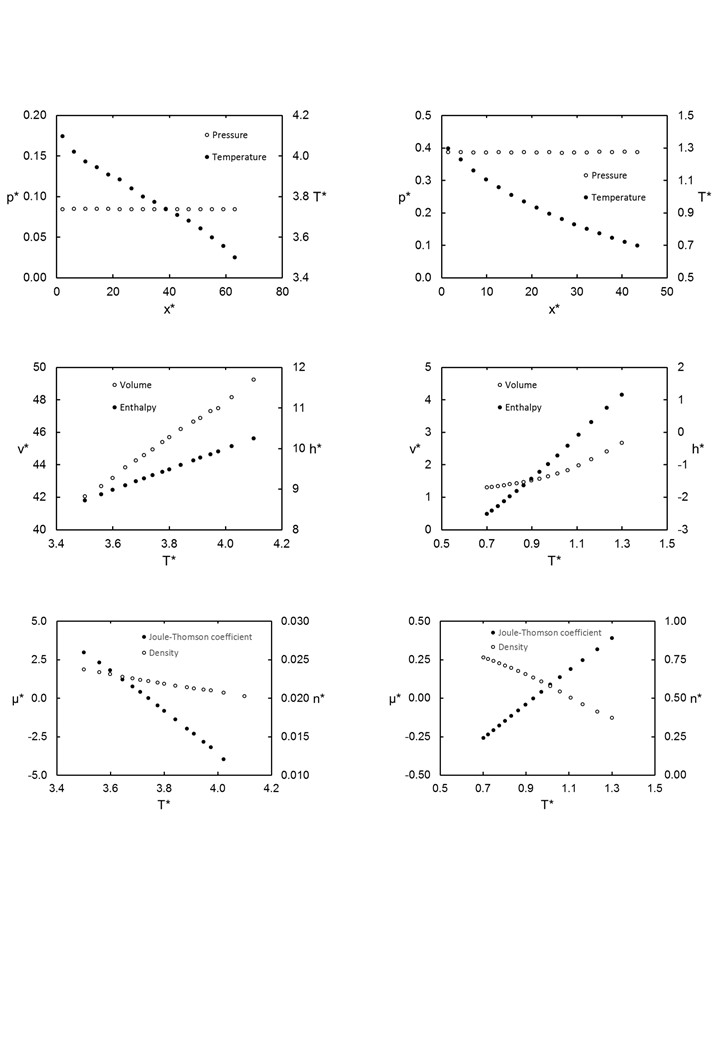}
\par\end{raggedright}
\caption{Examples of intermediate results from NEMD used to compute the Joule-Thomson coefficient at $n^*=0.022$ (left) and $n^*=0.62$ (right).}
\label{JTmd}
\end{figure}

The JT coefficient was computed as described in Sections \ref{JTnoneq} and \ref{isobaric}. Examples of intermediate results for the properties entering Eq. \eqref{eqn:JTevalb}, obtained with NEMD as described in Section \ref{JTnoneq}, are shown in Figure \ref{JTmd}. The data in the top panels were used to ascertain that the pressure was constant throughout the system with a temperature gradient. The data in the middle panels were used to compute the heat capacity (Eq. \eqref{eqn:heatcapacity}) and the thermal expansion coefficient (Eq. \eqref{eqn:thermalexpansion}). The data in the bottom panels were used to determine the temperature and density corresponding to $\mu_{\text{JT}}= constant$. The numerical values were computed from linear or second-order polynomial fits to the data.
Simulations were made for $constant \in \{-0.2,-0.1,0,0.1,0.2 \}$, but only the inversion curve ($constant=0$) will be shown here.

\begin{figure}[tbp]
\includegraphics[scale=0.6]{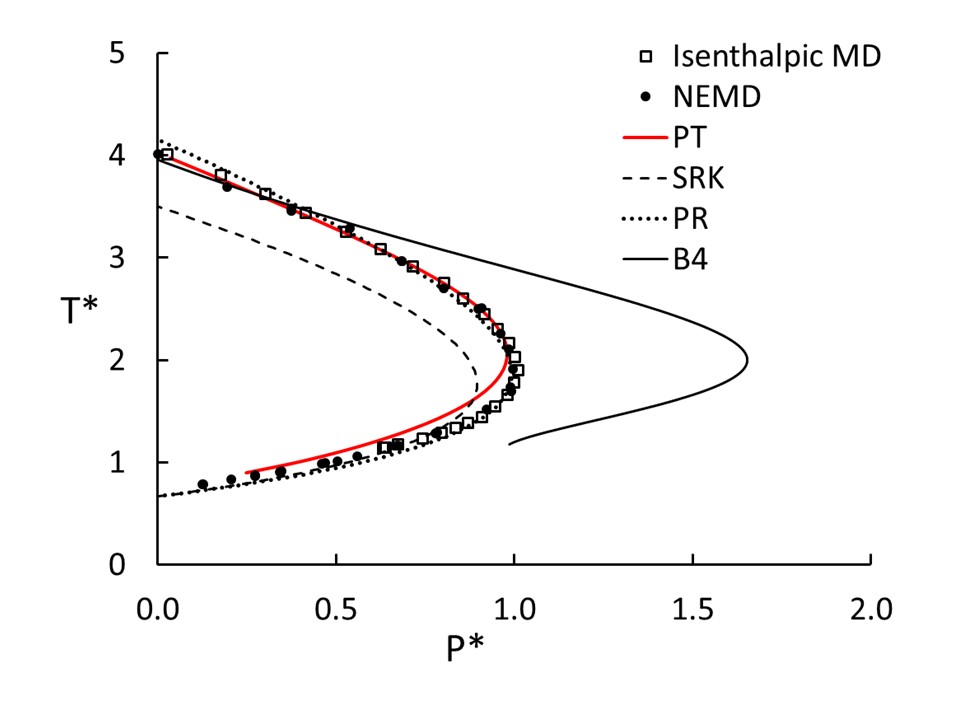}
\caption{Plots representing the Joule-Thomson inversion. The dots and squares are MD results. The lines represent the JT inversion curve as predicted by the second-order perturbation theory (PT), the virial expansion including $B_4$ (B4), and the two cubic EoS (SRK and PR).}
\label{fig:JTmdresults}
\end{figure}

Results from the NEMD and isenthalpic simulations for the JT inversion curve are shown as dots and squares, respectively, in Figure \ref{fig:JTmdresults}.

\subsubsection{The virial expansion}
\label{resultsJTvirial}

In the zero-density limit, the Joule-Thomson coefficient for a monatomic gas is given by
\begin{equation}
\mu_{\textnormal{JT}}=\frac{2T}{5k_{B}}\left[\frac{dB_{2}(T)}{dT}-\frac{B_{2}(T)}{T}\right].
\end{equation}
The numerical differentiation of $B_2(T)$ was done with a five-point difference method.
If the virial expansion is truncated at the $m^{\text{th}}$ term, applying Eq. \eqref{eqn:virialexp} to Eq. \eqref{eqn:JTeval} gives
\begin{equation}
\sum_{k=2}^m \left [ T B_k ' -(k-1)B_k \right ] n^{k-2} = 0,
\label{eqn:trunc}
\end{equation}
which is a polynomial equation in the density. Solutions of Eq. \eqref{eqn:trunc}  yield $\{T,n\}$ sets, which can be converted into $\{T,P\}$ using Eq. \eqref{eqn:virialexp}. The JT inversion curve computed from the virial expansion is compared with MD results in Figure \ref{fig:JTmdresults}. The density range spanned by the data is $0<n^*<0.7$ and the range of temperature is $1 < T^* < 4$, with a peak in $P^*$ at $P^* \approx 1.6$ and $T^* \approx 2$. 

The inversion curve predicted by virial coefficients shows good agreement with the other methods at low values of $P^*$ and high values of $T^*$, where the virial expansion is known to work well. The temperature corresponding to the maximum in $P^*$ is also in good agreement with the simulation results, but the maximum pressure is almost a factor 2 off, greatly overestimating the maximum $P^*$. This is surprising because the maximum corresponds to $n^* \approx 0.4$, \textit{i.e.} a density and temperature where both the compression factor and its derivative are in excellent agreement with the simulation results (see Figure \ref{fig:Z}). 

\subsubsection{Perturbation theory}
With the parameterization of the perturbation theory detailed in Section \ref{perturbation}, we used the thermodynamic framework presented in Ref.~\cite{Wilhelmsen2017} to compute the JT inversion curve as analytical derivatives of the parameterization.
The resulting JT inversion curve computed from the Barker-Henderson second-order perturbation theory is shown in Figure \ref{fig:JTmdresults}. The theory reproduces simulation data accurately at higher temperatures, but less accurately at lower temperatures (higher densities). This can be explained on the basis of the densities, which are gas-like in the top-part of Figure \ref{fig:JTmdresults}, but liquid-like in the bottom part. As shown in Figure \ref{fig:cubicsatdens}, the perturbation theory is more accurate for the vapour-phase than for the liquid-phase. 

\subsubsection{Cubic equations of state}
For the cubic EoS, the conclusions are similar as for the saturation densities, that PR reproduces more accurately the simulation results for the JT inversion curve than SRK, where PR gives an reasonably accurate representation of the simulation data.

\subsection{The speed of sound}
\label{sound}

The results for the speed of sound are given in Table \ref{table:sound}. All three methods reported here are within 5\% of the MD data for the gas state. For the supercritical state, the agreement is poor, between 15\% and 30\%. For the liquid state, the agreement is within 3\% for SRK and PT, but rather poor for PR. 

A closer analysis of the three contributions to the speed of sound, as given by Eq. \eqref{eqn:soundspeed}, shows that the failure of the SRK in the supercritical state is mostly due to the isothermal compressibility, which is off by more than a factor 2. On the other hand, the isothermal compressibility is off by some 40\% for PR in the liquid state. Like we noted for the virial expansion in Section \ref{resultsJTvirial}, this is surprising since Figure \ref{fig:cubicsuper} shows good agreement with MD data for the PR in the liquid region. Again, this illustrates the difficulties in finding a reliable thermodynamic model that can predict not only the first derivatives of the free energy (\textit{e.g.} the pressure), but also the second derivatives (\textit{e.g.} the compressiblity). The second-order perturbation theory actually does a fair job for all three states examined here.

\begin{table}
\centering
\caption{Speed of sound for three thermodynamic states.}
\vspace{0.4 cm}
\renewcommand*{\arraystretch}{1.3}
\begin{tabular}{c c c c c c c} \hline
State & $T^*$ & $n^*$ & MD & SRK & PR & PT \\[3pt]
\hline
Gas & $0.70$ & $0.02$  & $0.99$ & $1.03$ & $1.03$ & $1.04$ \\
Supercritical & $0.99$ & $0.40$  & $ 1.47 $ & $1.88$ & $1.69$ & $1.78$ \\
Liquid & $0.70$ & $0.80$ & $ 5.27 $ & $5.56$ & $3.64$ & $5.11$ \\
\hline
\end{tabular}
\label{table:sound}    
\end{table}

\subsection{Supercritical pressure-volume isotherms}
\begin{figure}
\includegraphics[scale=0.55]{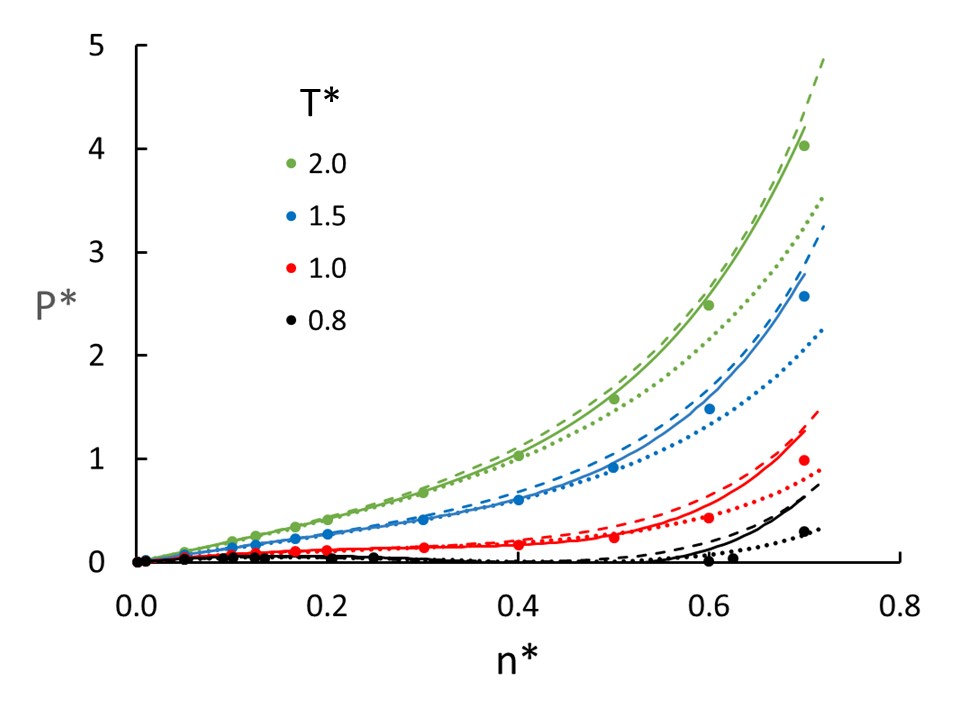}\caption{Isotherms from SRK (dashed lines), PR (dotted lines), second-order perturbation theory (solid lines), and MD simulations (dots).}
\label{fig:cubicsuper}
\end{figure}

 Figure~\ref{fig:cubicsuper} shows four $PV$ isotherms from PR (dotted lines), SRK (dashed liens) and the second order perturbation theory (solid lines) in comparison to MD simulation data. Whereas  all models show good agreement with simulation data at densities below $n^*<0.4$ (see also Fig. \ref{fig:Z}), with PR slightly better than the SRK, the situation is different at high densities. We find that PR is superior to SRK for $T^*<1.0$ and $n^*>0.6$, which supports the finding for liquid-phase density above, but for $T^*>1.5$, the SRK is superior. The perturbation theory gives results that are slightly better than SRK.

\section{Conclusions}
\label{conclusions}

In this work, we have presented a comprehensive map of the thermodynamic properties of the LJ/s model obtained with MD and GEMC simulations. We have estimated the critical point for the LJ/s model to be $T_\text{c}^*=0.885 \pm 0.002$, $P_\text{c}^*=0.075 \pm 0.001$, and $n_\text{c}^*=0.332 \pm 0.001$, respectively. The triple point was found to be $T_\text{tp}^*=0.547 \pm 0.005$ and $P_\text{tp}^*=0.0016 \pm 0.0002$, and the acentric factor was determined to be $0.07 \pm 0.02$. This enabled an assessment of the ability of the cubic equations of state, SRK and PR, to represent the thermodynamic properties of the LJ/s model. We further compared the simulation data to the virial expansion and a Barker-Henderson second-order perturbation theory.

For the gas/liquid coexistence densities, we found that PR was closer to the liquid-phase densities than SRK, while both EoS reproduced the vapour-phase densities and the saturation pressure within the accuracy of the simulations. Even though the second-order perturbation theory improved the first-order estimates for the critical temperature and pressure, the first-order perturbation theory gave liquid-phase densities that were closer to simulations. The reason for this was hypothesised to be that the mean compressibility approximation gives a poor representation at higher densities, and that a correction factor is needed. We found that the perturbation theory for the LJ/s model was inferior to that of the LJ model due to the shorter range of the potential.

We further presented surface tensions for the LJ/s model, gas/solid and liquid/solid coexistence densities, supercritical isotherms, speed of sound and Joule-Thomson inversion curves. All of the models and theories were in qualitative agreement with the simulation data, but none of them was capable of accurately reproducing the thermodynamic properties of the LJ/s model.
Our main conclusion is therefore that we at the moment do not have a theory or model that adequately represents the thermodynamic properties of the LJ/s system.

A suggestion for future work is to develop a more accurate perturbation theory that is capable of handling also short-ranged potentials such as the LJ/s model. 

\section*{Acknowledgements}
Morten Hammer and Øivind Wilhelmsen have been supported by the HYVA project, which is part of the Strategic Institute Programme of SINTEF Energy Research funded through the Basic Research Funding scheme of the Research Council of Norway. Computer resources were provided by Department of Chemistry at NTNU and by the HPC resources at UiT provided by NOTUR, http://www.sigma2.no. We are thankful to Olav Galteland for implementing the LJ/s potential in LAMMPS (http://folk.ntnu.no/olavgal/spline/).

\bibliographystyle{ieeetr}
\bibliography{LJsplineEOS}

\newpage

\large{Supplementary information}

\begin{table}[!ht]
\caption{Results from MD simulations.The computed properties are the number density ($n^*$), the volume per particle ($v^*=1/n^*$), pressure ($P^*$), potential energy ($U_\text{pot}/Nk_\text{B}T$), enthalpy per particle ($H^*/N$), and compression factor ($Z$). Data in the two-phase region have been omitted in the table.}
\vspace{0.1cm}
\scriptsize{
\begin{tabular}{p{0.5cm} c c c c c c}
\hline \\
$T^*$ & $n^*$ & $v^*$ & $P^*$ & $U_p^*/Nk_\text{B}T^*$ & $H^*/N$ & $Z$ \\
\\
\hline 
2.00 & 0.0010$\pm$0.0000 & 1000$\pm$4 & 0.0020$\pm$0.0000 & -0.0011$\pm$0.0001 & 4.9977$\pm$0.0003 & 1.000$\pm$0.006 \\
2.00 & 0.0100$\pm$0.0000 & 100.0$\pm$0.3 & 0.0200$\pm$0.0001 & -0.0222$\pm$0.0002 & 4.9525$\pm$0.0006 & 0.998$\pm$0.004 \\
2.00 & 0.0500$\pm$0.0001 & 19.99$\pm$0.06 & 0.0996$\pm$0.0003 & -0.1100$\pm$0.0004 & 4.772$\pm$0.001 & 0.996$\pm$0.004 \\
2.00 & 0.1000$\pm$0.0002 & 10.00$\pm$0.02 & 0.2000$\pm$0.0003 & -0.2177$\pm$0.0005 & 4.564$\pm$0.002 & 1.000$\pm$0.002 \\
2.00 & 0.1250$\pm$0.0002 & 8.00$\pm$0.01 & 0.2514$\pm$0.0004 & -0.2706$\pm$0.0006 & 4.470$\pm$0.002 & 1.005$\pm$0.002 \\
2.00 & 0.1666$\pm$0.0002 & 6.001$\pm$0.008 & 0.3401$\pm$0.0005 & -0.3583$\pm$0.0007 & 4.324$\pm$0.002 & 1.020$\pm$0.002 \\
2.00 & 0.2000$\pm$0.0003 & 4.999$\pm$0.008 & 0.4152$\pm$0.0007 & -0.4283$\pm$0.0007 & 4.219$\pm$0.002 & 1.038$\pm$0.002 \\
2.00 & 0.3000$\pm$0.0003 & 3.333$\pm$0.003 & 0.6761$\pm$0.0008 & -0.6360$\pm$0.0006 & 3.982$\pm$0.002 & 1.127$\pm$0.002 \\
2.00 & 0.4000$\pm$0.0002 & 2.500$\pm$0.002 & 1.034$\pm$0.001 & -0.8437$\pm$0.0005 & 3.898$\pm$0.002 & 1.293$\pm$0.001 \\
2.00 & 0.5001$\pm$0.0002 & 1.9998$\pm$0.0009 & 1.580$\pm$0.001 & -1.0521$\pm$0.0006 & 4.056$\pm$0.003 & 1.580$\pm$0.001 \\
2.00 & 0.6000$\pm$0.0002 & 1.6667$\pm$0.0006 & 2.483$\pm$0.002 & -1.2547$\pm$0.0006 & 4.629$\pm$0.002 & 2.069$\pm$0.002 \\
2.00 & 0.7000$\pm$0.0002 & 1.4286$\pm$0.0003 & 4.028$\pm$0.003 & -1.4345$\pm$0.0004 & 5.885$\pm$0.003 & 2.877$\pm$0.002 \\
\hline
1.50 & 0.0010$\pm$0.0000 & 995$\pm$6 & 0.0015$\pm$0.0000 & -0.0017$\pm$0.0001 & 3.7466$\pm$0.0002 & 1.000$\pm$0.009 \\
1.50 & 0.0100$\pm$0.0000 & 99.9$\pm$0.3 & 0.0149$\pm$0.0000 & -0.0342$\pm$0.0003 & 3.6845$\pm$0.0007 & 0.991$\pm$0.004 \\
1.50 & 0.0500$\pm$0.0002 & 20.00$\pm$0.06 & 0.0718$\pm$0.0002 & -0.1683$\pm$0.0006 & 3.433$\pm$0.001 & 0.957$\pm$0.004 \\
1.50 & 0.1000$\pm$0.0002 & 10.00$\pm$0.02 & 0.1387$\pm$0.0003 & -0.3302$\pm$0.0008 & 3.142$\pm$0.002 & 0.925$\pm$0.003 \\
1.50 & 0.1251$\pm$0.0002 & 8.00$\pm$0.01 & 0.1711$\pm$0.0003 & -0.4103$\pm$0.0009 & 3.003$\pm$0.002 & 0.912$\pm$0.002 \\
1.50 & 0.1667$\pm$0.0002 & 6.000$\pm$0.008 & 0.2246$\pm$0.0003 & -0.5389$\pm$0.0008 & 2.789$\pm$0.002 & 0.898$\pm$0.002 \\
1.50 & 0.2000$\pm$0.0003 & 5.000$\pm$0.008 & 0.2679$\pm$0.0004 & -0.640$\pm$0.001 & 2.630$\pm$0.002 & 0.893$\pm$0.002 \\
1.50 & 0.3000$\pm$0.0004 & 3.333$\pm$0.005 & 0.4109$\pm$0.0007 & -0.935$\pm$0.001 & 2.217$\pm$0.003 & 0.913$\pm$0.002 \\
1.50 & 0.4000$\pm$0.0004 & 2.500$\pm$0.003 & 0.6037$\pm$0.0009 & -1.227$\pm$0.001 & 1.919$\pm$0.002 & 1.006$\pm$0.002 \\
1.50 & 0.5000$\pm$0.0003 & 2.000$\pm$0.001 & 0.914$\pm$0.001 & -1.5223$\pm$0.0009 & 1.794$\pm$0.002 & 1.218$\pm$0.002 \\
1.50 & 0.6000$\pm$0.0003 & 1.6666$\pm$0.0007 & 1.483$\pm$0.002 & -1.822$\pm$0.001 & 1.988$\pm$0.002 & 1.647$\pm$0.002 \\
1.50 & 0.7000$\pm$0.0002 & 1.4286$\pm$0.0004 & 2.577$\pm$0.002 & -2.1081$\pm$0.0006 & 2.769$\pm$0.002 & 2.454$\pm$0.002 \\
\hline
1.20 & 0.0010$\pm$0.0000 & 999$\pm$5 & 0.0012$\pm$0.0000 & -0.0047$\pm$0.0002 & 2.9922$\pm$0.0004 & 0.998$\pm$0.008 \\
1.20 & 0.0100$\pm$0.0000 & 100.1$\pm$0.3 & 0.0118$\pm$0.0000 & -0.0488$\pm$0.0005 & 2.9192$\pm$0.0009 & 0.981$\pm$0.004 \\
1.20 & 0.0500$\pm$0.0001 & 19.99$\pm$0.05 & 0.0548$\pm$0.0001 & -0.2403$\pm$0.0009 & 2.607$\pm$0.002 & 0.913$\pm$0.003 \\
1.20 & 0.0999$\pm$0.0003 & 10.01$\pm$0.03 & 0.1008$\pm$0.0003 & -0.469$\pm$0.002 & 2.246$\pm$0.003 & 0.841$\pm$0.004 \\
1.20 & 0.1250$\pm$0.0003 & 8.00$\pm$0.02 & 0.1214$\pm$0.0002 & -0.579$\pm$0.002 & 2.076$\pm$0.003 & 0.809$\pm$0.003 \\
1.20 & 0.1669$\pm$0.0005 & 5.99$\pm$0.02 & 0.1531$\pm$0.0003 & -0.758$\pm$0.002 & 1.808$\pm$0.004 & 0.765$\pm$0.003 \\
1.20 & 0.1998$\pm$0.0006 & 5.00$\pm$0.01 & 0.1770$\pm$0.0004 & -0.892$\pm$0.003 & 1.615$\pm$0.005 & 0.738$\pm$0.003 \\
1.20 & 0.2999$\pm$0.0008 & 3.335$\pm$0.009 & 0.2487$\pm$0.0005 & -1.278$\pm$0.004 & 1.096$\pm$0.005 & 0.691$\pm$0.002 \\
1.20 & 0.3999$\pm$0.0007 & 2.500$\pm$0.004 & 0.3417$\pm$0.0007 & -1.646$\pm$0.003 & 0.679$\pm$0.003 & 0.712$\pm$0.002 \\
1.20 & 0.5001$\pm$0.0005 & 2.000$\pm$0.002 & 0.505$\pm$0.001 & -2.018$\pm$0.002 & 0.388$\pm$0.002 & 0.841$\pm$0.002 \\
1.20 & 0.6000$\pm$0.0003 & 1.6667$\pm$0.0008 & 0.853$\pm$0.001 & -2.405$\pm$0.001 & 0.336$\pm$0.002 & 1.185$\pm$0.002 \\
1.20 & 0.7000$\pm$0.0002 & 1.4285$\pm$0.0004 & 1.642$\pm$0.002 & -2.7957$\pm$0.0008 & 0.791$\pm$0.002 & 1.954$\pm$0.002 \\
\hline
1.00 & 0.0010$\pm$0.0000 & 998$\pm$6 & 0.0010$\pm$0.0000 & -0.0035$\pm$0.0002 & 2.4950$\pm$0.0003 & 0.999$\pm$0.008 \\
1.00 & 0.0100$\pm$0.0000 & 100.2$\pm$0.4 & 0.0097$\pm$0.0000 & -0.0661$\pm$0.0008 & 2.405$\pm$0.001 & 0.971$\pm$0.006 \\
1.00 & 0.0500$\pm$0.0002 & 20.00$\pm$0.07 & 0.0431$\pm$0.0001 & -0.331$\pm$0.002 & 2.031$\pm$0.003 & 0.862$\pm$0.004 \\
1.00 & 0.0999$\pm$0.0004 & 10.01$\pm$0.04 & 0.0745$\pm$0.0002 & -0.644$\pm$0.003 & 1.602$\pm$0.004 & 0.746$\pm$0.003 \\
1.00 & 0.1252$\pm$0.0006 & 7.99$\pm$0.04 & 0.0868$\pm$0.0003 & -0.797$\pm$0.004 & 1.396$\pm$0.005 & 0.693$\pm$0.004 \\
1.00 & 0.1667$\pm$0.0008 & 6.00$\pm$0.03 & 0.1032$\pm$0.0003 & -1.038$\pm$0.005 & 1.082$\pm$0.007 & 0.619$\pm$0.004 \\
1.00 & 0.200$\pm$0.001 & 5.00$\pm$0.03 & 0.1140$\pm$0.0004 & -1.215$\pm$0.008 & 0.86$\pm$0.01 & 0.571$\pm$0.004 \\
1.00 & 0.301$\pm$0.002 & 3.32$\pm$0.02 & 0.1403$\pm$0.0006 & -1.70$\pm$0.01 & 0.26$\pm$0.01 & 0.466$\pm$0.004 \\
1.00 & 0.400$\pm$0.002 & 2.50$\pm$0.01 & 0.1700$\pm$0.0007 & -2.129$\pm$0.008 & -0.204$\pm$0.009 & 0.425$\pm$0.003 \\
1.00 & 0.500$\pm$0.001 & 1.999$\pm$0.005 & 0.235$\pm$0.001 & -2.551$\pm$0.005 & -0.582$\pm$0.005 & 0.469$\pm$0.002 \\
1.00 & 0.6000$\pm$0.0004 & 1.667$\pm$0.001 & 0.425$\pm$0.001 & -3.007$\pm$0.002 & -0.799$\pm$0.001 & 0.708$\pm$0.002 \\
1.00 & 0.7000$\pm$0.0003 & 1.4286$\pm$0.0005 & 0.984$\pm$0.001 & -3.494$\pm$0.001 & -0.589$\pm$0.002 & 1.405$\pm$0.002 \\
\hline
0.95 & 0.0010$\pm$0.0000 & 999$\pm$6 & 0.0009$\pm$0.0000 & -0.0037$\pm$0.0002 & 2.3700$\pm$0.0003 & 0.998$\pm$0.008 \\
0.95 & 0.0100$\pm$0.0000 & 100.0$\pm$0.4 & 0.0092$\pm$0.0000 & -0.0739$\pm$0.0007 & 2.274$\pm$0.001 & 0.967$\pm$0.006 \\
0.95 & 0.0500$\pm$0.0001 & 20.00$\pm$0.05 & 0.0401$\pm$0.0001 & -0.364$\pm$0.002 & 1.882$\pm$0.002 & 0.845$\pm$0.003 \\
0.95 & 0.1000$\pm$0.0005 & 10.00$\pm$0.05 & 0.0676$\pm$0.0002 & -0.713$\pm$0.004 & 1.424$\pm$0.006 & 0.712$\pm$0.004 \\
0.95 & 0.1252$\pm$0.0009 & 7.99$\pm$0.06 & 0.0777$\pm$0.0003 & -0.883$\pm$0.007 & 1.206$\pm$0.009 & 0.653$\pm$0.005 \\
0.95 & 0.167$\pm$0.001 & 5.99$\pm$0.05 & 0.0903$\pm$0.0003 & -1.15$\pm$0.01 & 0.88$\pm$0.01 & 0.570$\pm$0.005 \\
0.95 & 0.199$\pm$0.002 & 5.02$\pm$0.04 & 0.0976$\pm$0.0003 & -1.34$\pm$0.01 & 0.64$\pm$0.01 & 0.516$\pm$0.004 \\
0.95 & 0.301$\pm$0.004 & 3.32$\pm$0.04 & 0.1129$\pm$0.0006 & -1.87$\pm$0.02 & 0.03$\pm$0.02 & 0.394$\pm$0.006 \\
0.95 & 0.400$\pm$0.003 & 2.50$\pm$0.02 & 0.1289$\pm$0.0007 & -2.31$\pm$0.01 & -0.44$\pm$0.01 & 0.339$\pm$0.003 \\
0.95 & 0.500$\pm$0.001 & 1.999$\pm$0.005 & 0.1689$\pm$0.0007 & -2.731$\pm$0.006 & -0.832$\pm$0.005 & 0.355$\pm$0.002 \\
0.95 & 0.6000$\pm$0.0005 & 1.667$\pm$0.001 & 0.318$\pm$0.001 & -3.201$\pm$0.003 & -1.086$\pm$0.002 & 0.557$\pm$0.002 \\
0.95 & 0.6999$\pm$0.0002 & 1.4288$\pm$0.0005 & 0.813$\pm$0.002 & -3.717$\pm$0.001 & -0.944$\pm$0.002 & 1.223$\pm$0.002 \\
\hline
\end{tabular}
}
\small{
\begin{tabular}{l}
 Continued next page \\
\hline
\end{tabular}
}
\label{table:isotherms1}
\end{table}

\begin{table}
\renewcommand\thetable{4 contd}
\caption{}
\vspace{0.1cm}
\scriptsize{
\begin{tabular}{c c c c c c c}
\hline \\
$T^*$ & $n^*$ & $v^*$ & $P^*$ & $U_p^*/Nk_\text{B}T^*$ & $H^*/N$ & $Z$ \\
\\
\hline
0.90 & 0.0010$\pm$0.0000 & 997$\pm$5 & 0.0009$\pm$0.0000 & -0.0041$\pm$0.0003 & 2.2447$\pm$0.0004 & 0.998$\pm$0.007 \\
0.90 & 0.0100$\pm$0.0000 & 100.1$\pm$0.4 & 0.0087$\pm$0.0000 & -0.081$\pm$0.001 & 2.144$\pm$0.001 & 0.963$\pm$0.005 \\
0.90 & 0.0500$\pm$0.0002 & 20.0$\pm$0.1 & 0.0371$\pm$0.0001 & -0.404$\pm$0.002 & 1.728$\pm$0.003 & 0.825$\pm$0.004 \\
0.90 & 0.0999$\pm$0.0006 & 10.0$\pm$0.1 & 0.0605$\pm$0.0002 & -0.796$\pm$0.006 & 1.239$\pm$0.008 & 0.673$\pm$0.005 \\
0.90 & 0.1247$\pm$0.0009 & 8.0$\pm$0.1 & 0.0681$\pm$0.0002 & -0.987$\pm$0.009 & 1.01$\pm$0.01 & 0.607$\pm$0.005 \\
0.90 & 0.167$\pm$0.002 & 6.0$\pm$0.1 & 0.0766$\pm$0.0003 & -1.29$\pm$0.02 & 0.65$\pm$0.02 & 0.511$\pm$0.006 \\
0.90 & 0.197$\pm$0.003 & 5.1$\pm$0.1 & 0.0805$\pm$0.0003 & -1.49$\pm$0.02 & 0.42$\pm$0.02 & 0.454$\pm$0.007 \\
0.90 & 0.299$\pm$0.009 & 3.3$\pm$0.1 & 0.0858$\pm$0.0004 & -2.07$\pm$0.04 & -0.23$\pm$0.05 & 0.32$\pm$0.01 \\
0.90 & 0.402$\pm$0.007 & 2.49$\pm$0.04 & 0.0893$\pm$0.0007 & -2.53$\pm$0.03 & -0.71$\pm$0.03 & 0.247$\pm$0.005 \\
0.90 & 0.501$\pm$0.002 & 1.998$\pm$0.008 & 0.1067$\pm$0.0008 & -2.945$\pm$0.009 & -1.087$\pm$0.008 & 0.237$\pm$0.002 \\
0.90 & 0.6001$\pm$0.0007 & 1.666$\pm$0.002 & 0.2117$\pm$0.0009 & -3.421$\pm$0.003 & -1.376$\pm$0.003 & 0.392$\pm$0.002 \\
0.90 & 0.7001$\pm$0.0003 & 1.4283$\pm$0.0005 & 0.6430$\pm$0.0016 & -3.967$\pm$0.001 & -1.302$\pm$0.002 & 1.020$\pm$0.003 \\
\hline
0.85 & 0.0010$\pm$0.0000 & 998$\pm$5 & 0.0009$\pm$0.0000 & -0.0043$\pm$0.0002 & 2.1196$\pm$0.0004 & 0.998$\pm$0.007 \\
0.85 & 0.0100$\pm$0.0000 & 100.0$\pm$0.4 & 0.0081$\pm$0.0000 & -0.091$\pm$0.001 & 2.012$\pm$0.001 & 0.958$\pm$0.005 \\
0.85 & 0.0500$\pm$0.0002 & 20.0$\pm$0.1 & 0.0340$\pm$0.0001 & -0.456$\pm$0.002 & 1.567$\pm$0.003 & 0.800$\pm$0.003 \\
0.85 & 0.1001$\pm$0.0008 & 10.0$\pm$0.1 & 0.0531$\pm$0.0002 & -0.91$\pm$0.01 & 1.03$\pm$0.01 & 0.624$\pm$0.005 \\
0.85 & 0.125$\pm$0.002 & 8.0$\pm$0.1 & 0.0581$\pm$0.0002 & -1.14$\pm$0.02 & 0.77$\pm$0.02 & 0.548$\pm$0.007 \\
$0.85^*$ & 0.166$\pm$0.004 & 6.0$\pm$0.2 & 0.0618$\pm$0.0003 & -1.52$\pm$0.04 & 0.36$\pm$0.04 & 0.44$\pm$0.01 \\
$0.85^*$ & 0.499$\pm$0.009 & 2.00$\pm$0.04 & 0.050$\pm$0.001 & -3.21$\pm$0.03 & -1.36$\pm$0.03 & 0.119$\pm$0.003 \\
$0.85^*$ & 0.600$\pm$0.001 & 1.665$\pm$0.003 & 0.108$\pm$0.001 & -3.674$\pm$0.005 & -1.668$\pm$0.004 & 0.212$\pm$0.002 \\
0.85 & 0.7000$\pm$0.0003 & 1.4286$\pm$0.0006 & 0.470$\pm$0.002 & -4.246$\pm$0.002 & -1.663$\pm$0.002 & 0.790$\pm$0.003 \\
\hline
0.80 & 0.0010$\pm$0.0000 & 999$\pm$6 & 0.0008$\pm$0.0000 & -0.0053$\pm$0.0003 & 1.9938$\pm$0.0005 & 0.997$\pm$0.009 \\
0.80 & 0.0100$\pm$0.0000 & 100.0$\pm$0.4 & 0.0076$\pm$0.0000 & -0.102$\pm$0.001 & 1.881$\pm$0.001 & 0.953$\pm$0.005 \\
$0.80^*$ & 0.0501$\pm$0.0002 & 19.97$\pm$0.09 & 0.0308$\pm$0.0001 & -0.524$\pm$0.004 & 1.397$\pm$0.005 & 0.769$\pm$0.004 \\
$0.80^*$ & 0.101$\pm$0.002 & 9.9$\pm$0.2 & 0.0448$\pm$0.0002 & -1.12$\pm$0.03 & 0.75$\pm$0.03 & 0.554$\pm$0.009 \\
$0.80^*$ & 0.12$\pm$0.02 & 8$\pm$1 & 0.0430$\pm$0.0003 & -1.6$\pm$0.3 & 0.3$\pm$0.3 & 0.43$\pm$0.06 \\
$0.80^*$ & 0.600$\pm$0.001 & 1.667$\pm$0.004 & 0.009$\pm$0.001 & -3.965$\pm$0.007 & -1.957$\pm$0.005 & 0.019$\pm$0.003 \\
0.80 & 0.7000$\pm$0.0003 & 1.4285$\pm$0.0005 & 0.295$\pm$0.002 & -4.562$\pm$0.001 & -2.027$\pm$0.002 & 0.528$\pm$ 0.003 \\
\hline
\label{table:isotherms2}
\end{tabular}
}
\\$ ^*$ Data points in the two-phase region, between the binodal and spinodal curves.

\end{table}

\begin{table}
\centering
\renewcommand\thetable{5}
\caption{Data for coexisting gas and liquid obtained with Gibbs ensemble Monte Carlo simulations.}
\vspace{0.3 cm}
\begin{tabular}{c c c c c c} \hline
$T^*$ & $P^*$ & $n_\text{gas}^*$ & $n_\text{liquid}^*$ & $(U/N)_\text{gas}^*$ & $(U/N)_\text{liquid}^*$ \\[2pt]
\hline
0.5500 & 0.0021 & 0.0040 & 0.8041 & 0.7730 & -3.6181 \\
0.6000 & 0.0046 & 0.0083 & 0.7780 & 0.8045 & -3.3484 \\
0.6250 & 0.0065 & 0.0115 & 0.7640 & 0.8118 & -3.2094 \\
0.6500 & 0.0089 & 0.0154 & 0.7490 & 0.8131 & -3.0673 \\
0.6750 & 0.0119 & 0.0205 & 0.7331 & 0.8043 & -2.9209 \\
0.7000 & 0.0157 & 0.0270 & 0.7158 & 0.7883 & -2.7695 \\
0.7250 & 0.0202 & 0.0351 & 0.6972 & 0.7586 & -2.6132 \\
0.7500 & 0.0256 & 0.0450 & 0.6760 & 0.7203 & -2.4466 \\
0.7750 & 0.0324 & 0.0589 & 0.6529 & 0.6505 & -2.2737 \\
0.8000 & 0.0399 & 0.0753 & 0.6250 & 0.5699 & -2.0828 \\
0.8250 & 0.0490 & 0.0996 & 0.5910 & 0.4383 & -1.8700 \\
0.8500 & 0.0597 & 0.1356 & 0.5449 & 0.2410 & -1.6154 \\
\hline
\end{tabular}
\end{table}

\begin{table}
\centering
\renewcommand\thetable{6}
\caption{Data for coexisting solid and liquid obtained with MD simulations as described in Section \ref{liquidsolid}}
\vspace{0.3 cm}
\begin{tabular}{c c c} \hline
$T^*$ & $n_\text{solid}^*$ & $n_\text{liquid}^*$ \\[2pt]
\hline
0.6000 & 0.9147 & 0.8280 \\
0.6500 & 0.9359 & 0.8487 \\
0.7000 & 0.9462 & 0.8618 \\
0.8000 & 0.9568 & 0.8842 \\
\hline
\end{tabular}
\end{table}

\begin{table}
\centering
\renewcommand\thetable{7}
\caption{Data for coexisting solid and gas obtained with MD simulations as described in Section \ref{gasliquid}}
\vspace{0.3 cm}
\begin{tabular}{c c c} \hline
$T^*$ & $n_\text{solid}^*$ & $n_\text{gas}^*$ \\[2pt]
\hline
0.4000 & 0.9238 & 0.0001 \\
0.4500 & 0.9063 & 0.0005 \\
0.5000 & 0.9138 & 0.0015 \\

\hline
\end{tabular}
\end{table}

\end{doublespacing}
\end{document}